\documentclass[aps,pra,twocolumn,superscriptaddress,10pt,showpacs]{revtex4-1}

\usepackage{graphicx}

\newcommand{\expect}[1]{{\left\langle{#1}\right\rangle}}
\newcommand{\tr}[2][]{\textnormal{tr}#1{{\left\{#2\right\}}}}
\newcommand{\ds}{\displaystyle}

\newcommand{\Eq}[2][Eq.~]{#1(\ref{eq:#2})}
\newcommand{\Fig}[2][Fig.~]{#1\ref{fig:#2}}

\newcommand{\half}{\frac{1}{2}}
\newcommand{\I}{\mathrm{i}}
\newcommand{\Exp}[1]{\mathrm{e}^{\mbox{\footnotesize$#1$}}}
\newcommand{\wf}[2][^{\ }]{\psi#1_{\mathrm{#2}}}
\newcommand{\mud}[2][^{\ }]{\phi#1_{\mathrm{#2}}}
\newcommand{\column}[2][c]{{\left(\begin{array}{#1}#2\end{array}\right)}}
\newcommand{\sqcol}[2][c]{{\left[\begin{array}{#1}#2\end{array}\right]}}
\newcommand{\adj}{^{\dagger}}
\newcommand{\tps}{^{\mathrm{T}}}
\newcommand{\repr}{\mathrel{\widehat{=}}}
\newcommand{\final}{\rho^{\ }_{\mathrm{fin}}}
\newcommand{\Real}[1]{\textnormal{Re}{\left(#1\right)}}
\newcommand{\BS}[1]{{\footnotesize BS#1}}
\newcommand{\Path}[1]{\textsc{#1}}
\newcommand{\sPath}[1]{\mbox{\footnotesize\textsc{#1}}}

\DeclareMathAlphabet{\vecfont}{OT1}{cmr}{bx}{it}
\renewcommand{\vec}[1]{\vecfont{#1}}

\begin{document}

\title{Past of a quantum particle: Common sense prevails}

\author{Berthold-Georg Englert}\email{cqtebg@nus.edu.sg}
\affiliation{Centre for Quantum Technologies, National University of %
  Singapore, 3 Science Drive 2, Singapore 117543, Singapore} 
\affiliation{Department of Physics, National University of Singapore, %
  2 Science Drive 3, Singapore 117542, Singapore} 
\affiliation{MajuLab, CNRS-UNS-NUS-NTU International Joint Unit, %
  UMI 3654, Singapore} 

\author{Kelvin Horia}\email{khoria@ntu.edu.sg}
\affiliation{Division of Physics and Applied Physics, %
             School of Physical and Mathematical Sciences, %
             Nanyang Technological University, %
             21 Nanyang Link, Singapore 637371, Singapore}

\author{Jibo Dai}
\altaffiliation{\makebox[\width][l]{Now at Data Storage Institute, A*STAR}}
\email{\makebox[\width][l]{dai\_jibo@dsi.a-star.edu.sg}}
\affiliation{Centre for Quantum Technologies, National University of %
  Singapore, 3 Science Drive 2, Singapore 117543, Singapore}

\author{Yink Loong Len}\email{yinkloong@quantumlah.org}
\affiliation{Centre for Quantum Technologies, National University of %
  Singapore, 3 Science Drive 2, Singapore 117543, Singapore} 

\author{Hui Khoon Ng}\email{cqtnhk@nus.edu.sg}
\affiliation{Yale-NUS College, 16 College Avenue West, Singapore 138527, %
  Singapore} 
\affiliation{Centre for Quantum Technologies, National University of %
  Singapore, 3 Science Drive 2, Singapore 117543, Singapore} 
\affiliation{MajuLab, CNRS-UNS-NUS-NTU International Joint Unit, %
  UMI 3654, Singapore} 

\date[]{Posted on the arXiv on 12 April 2017; updated on 27 August 2017}

\begin{abstract}
We analyze Vaidman's three-path interferometer with weak path marking
[\pra\ \textbf{87}, 052104 (2013)] and find that common sense yields correct
statements about the particle's path through the interferometer.
This disagrees with the original claim that the particles have discontinuous
trajectories at odds with common sense.  
In our analysis, ``the particle's path'' has operational meaning as acquired
by a path-discriminating measurement. 
For a quantum-mechanical experimental demonstration of the case, one should
perform a single-photon version of the experiment by Danan \textit{et al.\/}
[\prl\ \textbf{111}, 240402 (2013)] with unambiguous path discrimination. 
We present a detailed proposal for such an experiment.
\end{abstract}


\maketitle

\section{Introduction}\label{sec:intro}
Vaidman argues that one can meaningfully talk about the past of a quantum
particle --- specifically: which path it took through an interferometer --- by
analyzing the faint trace left along the path by weak, almost non-disturbing,
measurements within a formalism that uses forward and backward evolving
quantum states \cite{Vaidman:13a,Vaidman:14}.
He so arrives at conclusions that contradict common sense: 
The particle can have trajectories that are not continuous.
These assertions are confirmed, or so it seems, by an experiment that uses
periodic beam deflections at acoustic frequencies to mark the path in an
optical three-path interferometer \cite{Danan+3:13}.  
Various aspects of this matter have been debated 
\cite{Li+2:13,Vaidman:13b,Saldanha:14,Salih:15,Danan+3:15,%
      Svensson:14a,Huang+6:14,Wiesniak:14,Svensson:14b,%
      Bartkiewicz+5:15,Wu+6:15,Alonso+1:15,Potocek+1:15,Li+3:15,%
      Vaidman:16d,Vaidman:16a,Bartkiewicz+5:16,Hashmi+3:16,%
      Sokolovski:17a,Griffiths:16,Vaidman:16b,Vaidman:16c,Bula+6:16,%
      Ben-Israel+6:17,Roy+1:17,Nikolaev:17a,Vaidman:17a, Vaidman:17b,%
      Zhou+7:17,Nikolaev:17b,Griffiths:17,Sokolovski:17b}:
whether there is a need for the backward evolving state and the weak values of
the two-state formalism, and how to exploit them correctly;
whether the experiment can be described by classical optics or by standard
quantum-optical methods;
whether a consistent-histories description is more appropriate; 
whether a modified experiment is enlightening; and others.
The debate is still going on.

One particular aspect, however, has not yet received the attention it deserves,
\textsl{viz.} the crucial step of extracting unambiguous which-path information
from the faint traces left by an individual particle on its way through the
interferometer. 
This extraction gives operational and quantitative meaning to the otherwise
vague concept of ``knowledge about the past of the quantum particle.''
We acquire such specific knowledge about the path of a particle just detected
by a suitable measurement of the quantum degrees of freedom that are used to
mark the path. 
In this context, what we learn depends much on the question we ask by the
chosen measurement, and not all questions are equally relevant.
It turns out that common sense prevails if the right question is asked.

Our treatment is entirely within the standard formalism of quantum mechanics
and does not rely on the two-state formalism
\cite{Aharonov+2:64,Aharonov+1:90} employed by Vaidman. 
While we do not question the validity of the two-state formalism, we see no
particular advantage in using it; 
the standard formalism offers a transparent way for studying 
the properties of ensembles that are both pre-selected and post-selected. 

We set the stage by reviewing Vaidman's three-path interferometer in
Sec.~\ref{sec:3paths}, thereby introducing the conventions we use for labeling
the four beam splitters, the three paths through the interferometer,
and the five checkpoints along the paths.
Owing to the high symmetry of the setup, only one common-sense path is
available for the particles from the source to the detector.

We then note the description of the pre-selected and post-selected interfering
particles in terms of a forward and a backward propagating wave function.
Both wave functions are equally crucial in Vaidman's criterion for
establishing where the particle has been at intermediate times.
We state this criterion in Vaidman's words \cite{Vaidman:16b} and then
recall his narrative of the particles' history as it follows from
his interpretation of the two wave functions and the weak values associated
with the three paths.
In this narrative the particles propagate along all three paths, which
is at odds with the single-path story told by common sense \cite{Vaidman:13a}. 

Section \ref{sec:WPM} deals with the weak path marking by which a particle
leaves faint traces at the various checkpoints on its way from the source to
the detector.
We conclude that destructive interference suppresses the traces at two
checkpoints, which explains why a particle has a discontinuous trajectory in
Vaidman's narrative. 

Then, in Sec.~\ref{sec:WPK1}, we examine the faint traces left by the particle
just detected.
Unambiguous path knowledge --- with an operational and quantitative meaning
--- is available for a small fraction of the particles, and the path is
unknown for all others.
Upon noting that the probability amplitudes processed by the final beam
splitter are incoherent, we present an argument that the particles with
unknown path have, in fact, followed the common-sense path; in the limit of
ever fainter traces, these are all particles.
This conclusion is further supported by an examination of the subensembles of
unambiguously known paths or an utterly unknown path and also by the
considerations of Sec.~\ref{sec:WPK2} where we re-examine the faint traces by
another measurement.  

In Sec.~\ref{sec:2paths}, we take a close look at the two-path interferometer
in Vaidman's three-path setup.
We confirm that every particle detected at the exit for destructive
interference has a known path, and explain why this is consistent with
unknowable paths as the precondition for perfect interference, constructive or
destructive. 
Vaidman's narrative is at odds with common sense for the two-path
interferometer, too; in addition, the weak values for the paths have singular
imaginary parts. 

We propose a single-photon experiment for the two-path
interferometer with weak path marking in Sec.~\ref{sec:proposal}.
An account of the laboratory realization of this proposal will be reported
elsewhere~\cite{Len+3:16}.

Finally, in Sec.~\ref{sec:1photon-3paths} we propose a single-photon version
of the three-path interferometer experiment of Ref.~\cite{Danan+3:13}. 
This proposal has not been realized as yet.  
Once performed, it will demonstrate that common sense does prevail.  

\vspace*{-1.1\baselineskip}

\section{Vaidman's three-path interferometer}\label{sec:3paths}
\begin{figure}
  \centering
  \includegraphics{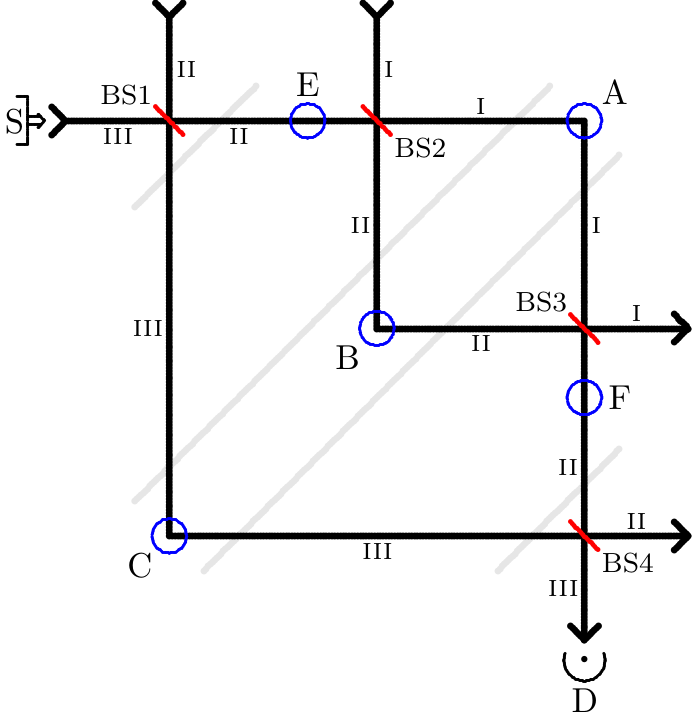}
  \caption{\label{fig:3paths}%
    Vaidman's three-path interferometer of Ref.~\cite{Vaidman:13a}.
    The quantum particle is emitted by source S, enters the interferometer
    at beam splitter \BS{1}, and is detected by detector D after
    exiting at beam splitter \BS{4}.
    On the way from S to D, the particle can take the path through 
    checkpoint C, or the paths through the internal Mach--Zehnder loop
    identified by checkpoints A and B.
    As a consequence of a weak coupling to the path-marker degrees of
    freedom, the particle leaves faint traces at these checkpoints, which
    enable the experimenter to infer the path actually followed.
    The additional checkpoints at E and F monitor passage into and out of the
    internal loop.
    The faint slanted lines connect simultaneous points on the three paths.
   }
\end{figure}

Vaidman's three-path interferometer of Ref.~\cite{Vaidman:13a} is depicted in
Fig.~\ref{fig:3paths} with the labeling conventions of
Ref.~\cite{Danan+3:13}. The particle (photon or otherwise) is emitted by source
S and detected by detector D, which it can reach along either one of the
three paths identified by the five checkpoints A, B, C, E, and F.
The setup is symmetric because the unitary three-by-three matrices for 
beam splitters \BS{1} and \BS{4} are the same as are the matrices for 
beam splitters \BS{2} and \BS{3} \cite{Horia:14},
\begin{eqnarray}\label{eq:2-1}
  U_1=U_4&=&\frac{1}{\sqrt{3}}\column[ccc]{\sqrt{3} & 0 & 0\\%
                                           0&-1&\sqrt{2}\\%
                                           0&\sqrt{2}&1}\,,
\nonumber\\ 
  U_2=U_3&=&\frac{1}{\sqrt{2}}\column[ccc]{1&1&0\\%
                                           -1&1&0\\%
                                           0&0&\sqrt{2}}\,.
\end{eqnarray}
They act on three-component columns of probability amplitudes associated with
the paths labeled \Path{i}, \Path{ii}, or \Path{iii} in~Fig.~\ref{fig:3paths}.

For instance, the column $\column[ccc]{0&0&1}\adj$ stands for the particle
emerging from source S and also for the state probed by detector D.
With no relative phases introduced in the various links inside the
interferometer, the probability that D detects the next particle emitted 
by S is
\begin{equation}\label{eq:2-2}
   \left|\column{0\\0\\1}\adj U_4U_3U_2U_1\column{0\\0\\1}\right|^2
  =\frac{1}{9}\,.
\end{equation}

\Fig[Figure~]{fwd-bwd} gives a more detailed account:
The thickness of the blue lines is proportional to the probability of finding
the particle there, should we look for it.
As a consequence of the symmetry of the setup, the particle has equal chance of
being found at checkpoints A, B, or C. 
We note that the Mach--Zehnder interferometer of the internal loop is balanced,
so that the particles do not pass checkpoint F.
Accordingly, common sense tells us that all particles detected by D followed
the path through checkpoint C and none of them came along EAF or EBF.

\begin{figure}
  \centering
  \includegraphics{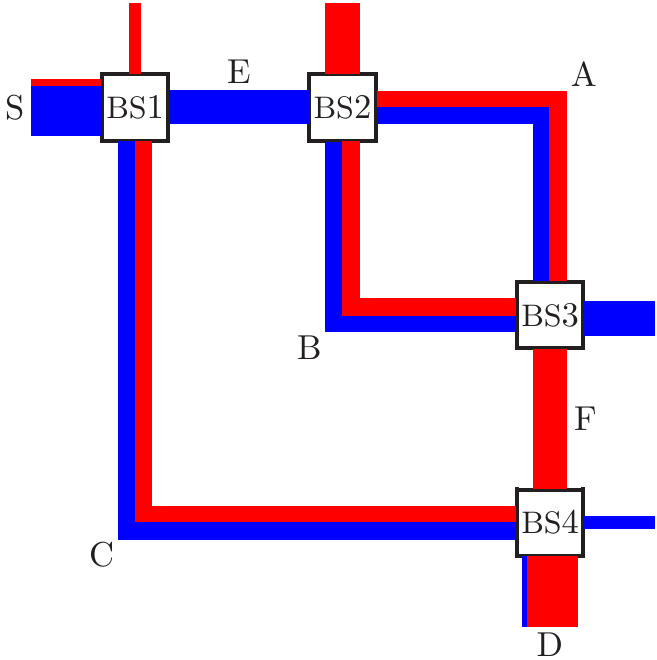}
  \caption{\label{fig:fwd-bwd}%
    Probabilities of finding the quantum particle between the source and the
    detector. 
    The thickness of the blue lines is proportional to the probability of
    finding the quantum particle there if we look for it.
    There are equal probabilities of $\frac{1}{3}$ at checkpoints A, B,
    and C, probability of $\frac{2}{3}$ at E, no probability at F, and the
    probability of reaching detector D is $\frac{1}{9}$.
    The red lines refer to the fictitious situation of preparing the quantum
    particle with non-zero amplitudes at all three entry ports such that it
    reaches D certainly.} 
\end{figure}

This common-sense conclusion is not shared by Vaidman \cite{Vaidman:13a}.
At the heart of his reasoning is a second state for which the relative
probabilities are indicated by the thickness of the red lines.
This fictitious state addresses the following question:
How would we need to prepare the particle so that it will be surely
detected by detector~D? 
Answer: We choose the probability amplitudes
\begin{equation}\label{eq:2-3}
  \left[\column{0\\0\\1}\adj U_4U_3U_2U_1\right]\adj
=\frac{1}{3}\column{-\sqrt{6}\\ \sqrt{2}\\ 1}\,.
\end{equation}
It is \emph{as if} the particle were injected into the interferometer by the
detector and propagating backward in time.
There is no mystery here, however.
The actual particle emerges from the source and the blue lines in
\Fig{fwd-bwd} apply.

Of course, the mathematical formalism offers the flexibility of evaluating
the probability amplitudes of \Eq{2-2} in various ways,
depending on which of the matrices in the product $U_4U_3U_2U_1$ act to the
right and which act to the left.
For each such split, we can find the so-called weak values of the projectors
onto the three paths, as illustrated by 
\begin{equation}\label{eq:2-4}
  {\left(\frac{1}{3}\right)}^{-1}
  \column{0\\0\\1}\adj U_4U_3\column{1\\0\\0}\column{1\\0\\0}\adj
   U_2U_1\column{0\\0\\1}=-1
\end{equation}
for path~\Path{i} in the symmetric split.
The normalizing prefactor divides by the amplitude that is squared in
\Eq{2-2}. 
By definition, then, the weak values for the three paths have unit sum for each
split.

\begin{table}
  \centering
  \caption{\label{tbl:weak}%
  The weak values of the projectors on the three paths for different ways of
  splitting the product $U_4U_3U_2U_1$ into operators acting to the right and
  to the left.
  The projectors are inserted at the split, indicted by the vertical line~$|$. 
  All $U$s act to the right in the first row, all act to the left in the last
  row.}
  \begin{tabular}{c@{\quad}c@{\quad}c@{\quad}c@{\enskip\ }}\hline\hline
  &&\makebox[0pt][c]{weak values}&\\
  split & \Path{i} & \Path{ii} & \Path{iii} \\  \hline
  $\big|U_4U_3U_2U_1$ & $0$ & $0$ & $1$\\
  $U_4\big|U_3U_2U_1$ & $0$ & $0$ & $1$\\
  $U_4U_3\big|U_2U_1$ & $-1$ & $1$ & $1$\\
  $U_4U_3U_2\big|U_1$ & $0$ & $0$ & $1$\\
  $U_4U_3U_2U_1\big|$ & $0$ & $0$ & $1$\\
  \hline\hline
  \end{tabular}
\end{table}

The three weak values for all five splits are reported in
Table~\ref{tbl:weak}. 
We observe that all three weak values have unit magnitude for the symmetric
split of the third row, which refers to projecting on the paths at 
checkpoints A, B, and C.
For all other splits, the weak values for paths~\Path{i} and~\Path{ii} vanish.

Consistent with this observation, we have coexisting blue and red lines in
Fig.~\ref{fig:fwd-bwd}, of equal thickness even, on the path along  
checkpoint C and also inside the internal loop. 
In Vaidman's view, the particle could only have been in these regions where we
have both the blue and the red probabilities \cite{Vaidman:16b}:
\begin{equation}\label{eq:2-5}
    \parbox{0.82\columnwidth}{The particle was present in paths of the
      interferometer in which there is an overlap of the forward and backward
      evolving wave functions.}
\end{equation}
In his narrative of the particle's history on the way from the source to the
detector, then, the particle was inside the internal loop at intermediate
times, but it did not pass  checkpoint E (no red probability) nor 
checkpoint F (no blue probability): 
The particle was inside the loop but it did not enter or leave.
At a certain instant, the particle was at checkpoints A \emph{and} B
\emph{and} C simultaneously \cite{weak1} (and could have left a weak
trace at all three checkpoints). 

Yet, we surely find the one particle traversing the interferometer in one
place only whenever we look for it. 
Nobody is looking, however.
It is impossible to reconcile this narrative with common sense.

\begin{figure}
  \centering
  \includegraphics{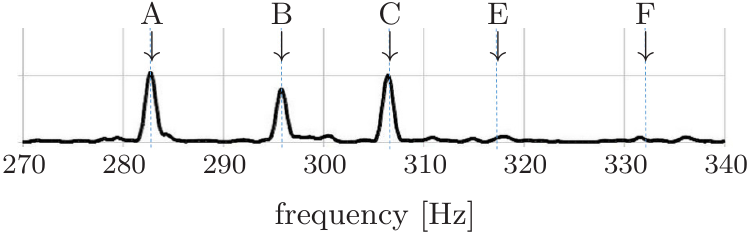}
  \caption{\label{fig:ExpData}%
   Data collected in the experiment by Danan \textit{et al.\/}~%
   \cite{Danan+3:13}.
   In this optical version of Vaidman's three-path interferometer, 
   light propagates through the interferometer in a horizontal plane.
   Path information is encoded by slightly deflecting the light beams out of
   this plane.
   This is achieved by mirrors (not drawn in \Fig{3paths}) at the checkpoints
   that oscillate about a horizontal axis with a small amplitude.
   The resulting upward or downward displacement at the detector 
   ($\sim600\,\mathrm{nm}$) is quite small compared with the transverse
   coherence length of the light ($\sim1.2\,\mathrm{mm}$) in order to ensure
   the weakness of the path marking. 
   The detector registers separately the light intensities above and below the
   horizontal plane, and the squared Fourier transform of their difference is
   the reported power spectrum, reproduced here from Figs.~2(b) and 3 in
   \cite{Danan+3:13}.
   The five mirrors at the five checkpoints oscillate with different frequencies
   between $280\,\mathrm{Hz}$ and $335\,\mathrm{Hz}$, so that the five
   well-separated peaks in the power spectrum can be associated with individual
   mirrors. 
   The peaks for checkpoints A, B, and C are of about equal height and much
   higher than those for checkpoints E and F, which are buried in the noise.}
\end{figure}

One could shrug and leave it at that.
As outrageous as they may be, we cannot test statements about the particle's
whereabouts unless we observe it on its way from the source to the
detector.
This observation has to be gentle on the particle in order to
not disturb the delicate balance of the interferometer. 
Something of this kind is accomplished by the weak path marking in the
optical experiment reported by Danan \textit{et al.\/} in
Ref.~\cite{Danan+3:13}. 
And, yes, the data show equally strong traces from checkpoints A, B, and
C but no traces from checkpoints E and F; 
see \Fig{ExpData}, where we reproduce the power spectrum of Figs.~2(b)
and 3 in \cite{Danan+3:13}.

Yes, these data are consistent with Vaidman's narrative, but does the
experiment really demonstrate the case? 
No, it doesn't, for at least two reasons.
First, the experiment is performed with classical light intensities, and
one does not need quantum mechanics for a comparison with theoretical
predictions; Maxwell's electrodynamics is quite sufficient.
Although one could invoke that, for such linear-optics circumstances, there is
a one-to-one correspondence between light intensities and photon probabilities
\cite{note:PowerSpectrum}, it remains true that the experiment by Danan
\textit{et al.\/} does not make any information available about individual
photons. 

Second, the data  are perfectly consistent with an alternative story: 
Each photon of a small fraction leaves a discernible trace at
checkpoint A \emph{or} at B \emph{or} at C, while most photons leave no trace
at all.  
This interpretation of the data, which is as natural as that offered
in \cite{Danan+3:13} if not more so, does not support Vaidman's criterion 
\Eq[]{2-5}.
(As discussed in Sec.~\ref{sec:WPM}, the peaks for checkpoints E and F are
suppressed by destructive interference.)

On the basis of the data reported by Danan \textit{et al.\/}, one cannot tell
whether Vaidman's narrative or the alternative story is correct.
A single-photon version of the experiment is needed to decide the matter.
Zhou \textit{et al.\/} \cite{Zhou+7:17} do send single photons through the
three-path interferometer, but their detection method is unable to extract all
the available path information, and as such is little better than the original
experiment in resolving the narrative. 
We present a proposal for a single-photon experiment that achieves full
path-information extraction in Sec.~\ref{sec:1photon-3paths}.

Such a single-photon experiment allows for
a meaningful statement about the path taken by the photon just detected.
The data of the classical-light experiment are good enough for ensemble
averages (expectation values).
Whether an individual photon leaves a trace at A, at B, or at C,
however, can only be inferred in a single-photon experiment.  
Our analysis concerns this single-photon situation, and the weak values of
Table~\ref{tbl:weak} play no role in it.
We predict that the data of an experiment with single photons will speak
against Vaidman's narrative and in favor of the alternative story.

\section{Weak path marking}\label{sec:WPM}
The oscillating-mirror method of imprinting a path mark on the light 
in the experiment of Danan \textit{et al.\/} \cite{Danan+3:13},
see the caption to \Fig{ExpData}, does not lend itself to
extraction of path knowledge about an individual photon, should one execute
the experiment in a single-photon fashion.
Statements of the kind ``the particle (= photon) is inside the internal loop
after passing checkpoint E and before reaching checkpoint F'' take for granted
that the photon is sufficiently localized --- having a longitudinal extension
of $3\,\mathrm{cm}$, say, roughly one-tenth of the distance between beam
splitters \BS{2} and \BS{3} (an example of Vaidman's ``localized wave
packet'' \cite{Vaidman:13a}).
Then, the photon takes $\sim0.1\,\mathrm{ns}$ to pass a point on its trajectory.
The mirrors, which have oscillation periods of several milliseconds, are
standing still for lapses of time so short.
None of the frequencies of the mirror oscillations is imprinted on the photon.
An individual photon is simply deflected up or down a bit, either by the
mirror at checkpoint C or jointly by the mirrors at E, A or B, and F.

We cannot infer the path taken from the observed deflection, as we cannot
tell which mirror(s) deflected the photon \cite{fn:EOM1}.
Therefore, we do not examine the experiment of Danan \textit{et al.\/}
\cite{Danan+3:13} in further detail, and rather exploit entanglement of a
more immediately useful kind for the path marking and the extraction of path
information. 

For the path marking, we use weak interactions at the checkpoints, described
by unitary operators $A$, $B$, $C$, $E$, and $F$ that differ slightly from the
identity operator and act on certain quantum degrees of freedom other than the
path degree of the interferometer.
These marker degrees of freedom are initially not correlated with the
particle's path degree of freedom, and we write $\rho$ for the statistical
operator of the initial path-marker state.

Then, the very small probability $\epsilon$ that the next particle reaches
checkpoint F is 
\begin{equation}\label{eq:3-1}
  \epsilon=\tr{T_{\mathrm{F}}^{\ }\rho T_{\mathrm{F}}\adj}
          =\expect{T_{\mathrm{F}}\adj T_{\mathrm{F}}^{\ }}\,,
\end{equation}
where the trace is over the path-marker degrees of freedom and 
\begin{eqnarray}\label{eq:3-2}
  T_{\mathrm{F}}^{\ }&=&\column{0\\1\\0}\adj U_3
                  \column[ccc]{A&0&0\\0&B&0\\0&0&C}U_2
                  \column[ccc]{1&0&0\\0&E&0\\0&0&1}U_1
                  \column{0\\0\\1}
                  \nonumber\\&=&\frac{1}{\sqrt{6}}(B-A)E
\end{eqnarray}
so that
\begin{equation}\label{eq:3-3}
  \epsilon=\frac{1}{6}\expect{E\adj(B-A)\adj(B-A)E}\,.
\end{equation}
We have $0<\epsilon\ll1$ under the weak path-marking circumstances of
interest, when the particle leaves but a faint trace of its path through the
interferometer. 

The unnormalized final state of the path marker, conditioned on the particle
detection by D, is $T_{\mathrm{fin}}^{\ }\rho T_{\mathrm{fin}}\adj$ with
\begin{eqnarray}\label{eq:3-4}
  T_{\mathrm{fin}}^{\ }&=&\column{0\\0\\1}\adj 
                    U_4\column[ccc]{1&0&0\\0&F&0\\0&0&1}U_3
                  \column[ccc]{A&0&0\\0&B&0\\0&0&C}U_2
   \\\nonumber
        &&\times
                  \column[ccc]{1&0&0\\0&E&0\\0&0&1}U_1
                  \column{0\\0\\1}
   =\frac{1}{3}\bigl[C+F(B-A)E\bigr]\,
\end{eqnarray}
where the coefficients $+1$, $+1$, and $-1$ for $C$, $FBE$, and $FAE$,
respectively, are reminiscent of the weak values in Table~\ref{tbl:weak} but
do not have the weak-value meaning.
The probability of detecting the particle is
\begin{eqnarray}\label{eq:3-5}
  &&\tr{T_{\mathrm{fin}}^{\ }\rho T_{\mathrm{fin}}\adj}
    =\expect{T_{\mathrm{fin}}\adj T_{\mathrm{fin}}^{\ }}\nonumber\\
&=&\frac{1}{9}\expect{\bigl[C+F(B-A)E\bigr]\adj\bigl[C+F(B-A)E\bigr]}\,,
\end{eqnarray}
and the normalized final state of the path marker is
\begin{eqnarray}\label{eq:3-6}
  \final&=&\frac{T_{\mathrm{fin}}^{\ }\rho T_{\mathrm{fin}}\adj}
              {\expect{T_{\mathrm{fin}}\adj T_{\mathrm{fin}}^{\ }}}
\\\nonumber  &=&
 \frac{\bigl[C+F(B-A)E\bigr]\rho\bigl[C+F(B-A)E\bigr]\adj}
 {\expect{\bigl[C+F(B-A)E\bigr]\adj\bigl[C+F(B-A)E\bigr]}}\,.
\end{eqnarray}
Since we want to investigate the faint traces left at individual checkpoints,
we shall now take for granted that each checkpoint has its own path-marker
degree of freedom and that there are no initial correlations between the marker
degrees of freedom \cite{fn:commute}.
Accordingly, $A$, $B$, $C$, $E$, and $F$ commute with one another and $\rho$
factorizes, 
\begin{equation}\label{eq:3-7}
  \rho=\rho^{\ }_{\mathrm{A}}\rho^{\ }_{\mathrm{B}}\rho^{\ }_{\mathrm{C}}
       \rho^{\ }_{\mathrm{E}}\rho^{\ }_{\mathrm{F}}\,.
\end{equation}
Then we have
\begin{equation}\label{eq:3-8}
  \epsilon=\frac{1}{6}\expect{(B-A)\adj(B-A)}
          =\frac{1}{3}-\frac{1}{3}\Real{\expect{A}^*\expect{B}}
\end{equation}
and
\begin{equation}\label{eq:3-9}
  \expect{T_{\mathrm{fin}}\adj T_{\mathrm{fin}}^{\ }}
   =\frac{1+6\epsilon}{9}-\frac{2}{9}
    \Real{\vphantom{\Big|}%
    \bigl(\expect{A}-\expect{B}\bigr)\expect{C}^*\expect{E}\expect{F}}
\end{equation}
where $\expect{X}=\tr[^{\ }_\mathrm{X}]{X\rho^{\ }_{\mathrm{X}}}$
for $X=A,B,C,E,F$.
We recover ${\epsilon=0}$ and 
$\expect{T_{\mathrm{fin}}\adj T_{\mathrm{fin}}^{\ }}=\frac{1}{9}$ for
${A=B=1}$, as we should.

The final statistical operator for checkpoint E is obtained by tracing
$\final$ over the other checkpoints, with the outcome
\begin{eqnarray}\label{eq:3-10}
  \rho^{\ }_{\mathrm{E,fin}}
&=&\frac{1}{9\expect{T_{\mathrm{fin}}\adj T_{\mathrm{fin}}^{\ }}}
  \Bigl[\rho^{\ }_{\mathrm{E}}+6\epsilon E\rho^{\ }_{\mathrm{E}}E\adj
\\\nonumber&&\hspace*{5em}\mbox{}        
   -2\Real{\vphantom{\Big|}%
           E\rho^{\ }_{\mathrm{E}}\bigl(\expect{A}-\expect{B}\bigr)
                \expect{C}^*\expect{F}}\Bigr]\,,
\end{eqnarray}
where $\Real{X}=\half(X+X\adj)$ for operator $X$,
and there is an analogous expression for $ \rho^{\ }_{\mathrm{F,fin}}$.
The final statistical operator for checkpoint C is 
\begin{eqnarray}\label{eq:3-11}
  \rho^{\ }_{\mathrm{C,fin}}
&=&\frac{1}{9\expect{T_{\mathrm{fin}}\adj T_{\mathrm{fin}}^{\ }}}
  \Bigl[C\rho^{\ }_{\mathrm{C}}C\adj+6\epsilon \rho^{\ }_{\mathrm{C}}
\\\nonumber&&\hspace*{4.5em}\mbox{}        
     -2\Real{\vphantom{\Big|}\rho^{\ }_{\mathrm{C}}C\adj
             \bigl(\expect{A}-\expect{B}\bigr)
                \expect{E}\expect{F}}\Bigr].
\end{eqnarray}
We note that $6\epsilon$ multiplies the initial state 
$\rho^{\ }_{\mathrm{C}}$ here whereas this factor multiplies the transformed 
$E\rho^{\ }_{\mathrm{E}}E\adj$ in $\rho^{\ }_{\mathrm{E,fin}}$.

The symmetry of the interferometer should be affected minimally by the weak
measurement.
In particular, we do not want to unbalance the internal Mach--Zehnder loop more
than is unavoidable.
Therefore, we require $\expect{A}=\expect{B}$, with their values determined by
\Eq{3-8}, 
\begin{equation}\label{eq:3-12}
  \expect{A}=\expect{B}=\sqrt{1-3\epsilon}
\end{equation}
and so arrive at
\begin{equation}\label{eq:3-13}
  \expect{T_{\mathrm{fin}}\adj T_{\mathrm{fin}}^{\ }}
=\frac{1+6\epsilon}{9}
\end{equation}
as well as
\begin{equation}\label{eq:3-14}
  \rho^{\ }_{\mathrm{E,fin}}
  =\frac{\rho^{\ }_{\mathrm{E}}+6\epsilon E\rho^{\ }_{\mathrm{E}}E\adj}
        {1+6\epsilon}
\end{equation}
and
\begin{equation}\label{eq:3-15}
  \rho^{\ }_{\mathrm{C,fin}}
   =\frac{C\rho^{\ }_{\mathrm{C}}C\adj+6\epsilon \rho^{\ }_{\mathrm{C}}}
        {1+6\epsilon}\,.
\end{equation}
Since ${\epsilon\ll1}$, we have 
$\rho^{\ }_{\mathrm{C,fin}}\simeq C\rho^{\ }_{\mathrm{C}}C\adj$
and
$\rho^{\ }_{\mathrm{E,fin}}\simeq\rho^{\ }_{\mathrm{E}}$, consistent with the
observation in the experiment of Ref.~\cite{Danan+3:13}: 
The particle leaves a trace at checkpoint C but not at E.

To shine some light on this matter, let us consider what happens when only $E$
is a genuine path-marking operator while $A$, $B$, $C$, and $F$ do not
entangle and just introduce phase factors $\Exp{\I\alpha}$, $\Exp{\I\beta}$,
$\Exp{\I\gamma}$, and $\Exp{\I\phi}$. 
Then the final state of \Eq{3-6} is
\begin{equation}\label{eq:3-16}
  \rho^{\ }_{\mathrm{fin}}\!\propto\!
       {\left[\Exp{\I\phi}{\left(\Exp{\I\alpha}-\Exp{\I\beta}\right)}E
              -\Exp{\I\gamma}\right]}\rho
       {\left[\Exp{\I\phi}{\left(\Exp{\I\alpha}-\Exp{\I\beta}\right)}E
              -\Exp{\I\gamma}\right]}\adj
\end{equation}
where we leave the normalization implicit.
The strength with which $E$ acts is determined by the probability amplitudes
$\Exp{\I\phi}\Exp{\I\alpha}$ and $\Exp{\I\phi}\Exp{\I\beta}$ for the
paths associated with checkpoints A and F or B and F, respectively.
Their difference is the net amplitude
\begin{equation}\label{eq:3-17}
  \Exp{\I\phi}\left(\Exp{\I\alpha}-\Exp{\I\beta}\right)
=2\I\Exp{\I[\phi+\frac{1}{2}(\alpha+\beta)]}\sin\frac{\alpha-\beta}{2}\,,
\end{equation}
which vanishes if we have perfect destructive interference at beam splitter
\BS{3} for the particles on the way to \BS{4}.
If there is a nonzero relative phase $\frac{1}{2}(\alpha-\beta)$ in the
internal loop, operator $E$ does act and the particle leaves a trace at
checkpoint E.

This solves the ``mystery of the missing trace:'' The destructive interference
that makes it hard for the particle to reach checkpoint F makes it equally
hard to leave a trace at checkpoint E. 
It is clear, then, that the absence of a trace at checkpoint E does not
indicate that the particle did not get there.
Rather, it was not given a chance to leave a trace of its passage
\cite{fn:EOM2}. 
A similar observation (for ${\epsilon=\frac{1}{3}}$) was made by Bartkiewicz
\textit{et al.\/} \cite{Bartkiewicz+5:15}.  

There are various manifestations of destructive interference in the three-path
interferometer of \Fig{3paths}. 
The lack of blue probability in \Fig{fwd-bwd} between beam splitters \BS{3}
and \BS{4} is one, the lack of red probability between \BS{1} and \BS{2} is
another.  
We regard the latter as destructive interference of the red amplitudes at
\BS{1}; Vaidman would view it as destructive interference of
backward-traveling red amplitudes at \BS{2}.
From either perspective, there is no mystery unless we choose to mystify the
familiar phenomenon of destructive interference.   

Accordingly, we improve on Vaidman's statement that the particle was inside
the internal loop but did not enter or leave:
The particle was inside but did not leave traces when entering and leaving.

\section{Which-path knowledge}\label{sec:WPK1}
\subsection{Unambiguous path discrimination}
\label{sec:WPK1a}
In expressions such as $\final$ in \Eq{3-6}, operators $E$ and $F$ always
appear together with $A$ or $B$.
No useful information (if any) can be extracted from the path markers at
checkpoints E and F that is not already made available at A and B. 
Therefore, we put $E=F=1$ and shall only work with $A$, $B$, and $C$.

\begin{figure}
  \centering
  \includegraphics{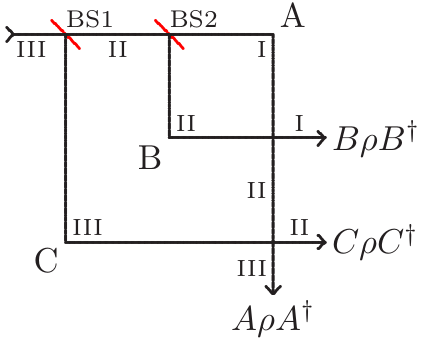}
  \caption{\label{fig:p-mode.a}%
    Marking the path through the interferometer.
    Unitary operators $A$, $B$, and $C$ are acting on the path-marker
    degrees of freedom when the particle passes through the respective
    checkpoints A, B, or C. 
    Here, beam splitters \BS{3} and \BS{4} are removed, and the conditional
    states of the path marker are indicated at the exit ports: $A\rho A\adj$,
    $B\rho B\adj$, and  $C\rho C\adj$ for the paths through checkpoint A,
    B, or C, respectively, where $\rho$ is the initial state of the path
    marker.  
  }
\end{figure}

We can then verify that the path marking works alright by removing beam
splitters \BS{3} and \BS{4} and using three detectors, one at each output
port; see \Fig{p-mode.a}.
The matrices for the unitary operators $U_3$ and $U_4$ of \Eq{2-1} are here
replaced by 
\begin{equation}\label{eq:4-0}
  U_3=\column[ccc]{0&1&0 \\1&0&0 \\ 0&0&1}\,,\quad
  U_4=\column[ccc]{1&0&0 \\0&0&1 \\ 0&1&0}\,.
\end{equation}
Just before the particle is detected, the joint state of the particle and the
path marker is
\begin{equation}\label{eq:4-1}
  \frac{1}{3}\column{B \\ C \\ A}\rho\column{B \\ C \\ A}\adj\,.
\end{equation}
The diagonal elements of the three-by-three matrix for the path degree of
freedom --- these are $\frac{1}{3}B\rho B\adj$, $\frac{1}{3}C\rho C\adj$, and
$\frac{1}{3}A\rho A\adj$ --- are the unnormalized final path-marker states
conditioned on detecting the particle at exits \Path{i}, \Path{ii}, or
\Path{iii}, respectively, after passing the corresponding checkpoint B, C, or A.
The traces of these diagonal elements are the respective probabilities, each
equal to one-third.
We acquire which-path knowledge by a measurement on the path-marker degrees of
freedom that distinguishes between the states $A\rho A\adj$,  $B\rho B\adj$,
and $C\rho C\adj$, which label the corresponding exit ports in \Fig{p-mode.a}.

It is important that there is an operational confirmation of the which-path
knowledge thus gained.
For this purpose, one experimenter (Alice) takes note which of the three
detectors found the particle, and a second experimenter (Bob) examines the
path marker to find out which checkpoint was visited.
Bob has various strategies at his disposal, among them two particularly
important ones.

In the first strategy, Bob makes an educated guess about each particle's path
and maximizes his chance of guessing right. 
For this purpose, he employs the so-called \emph{minimum-error measurement}
\cite{Helstrom:76}, and will guess right as often as possible without,
however, being sure about even one particle's path.
In the second strategy, he wants to be absolutely sure about the paths of some
particles at the expense of having no clue which path was taken by the other
particles, and chooses his measurement such that the fraction of particles with
surely-known paths is as large as possible. 
The \emph{measurement for unambiguous discrimination} 
\cite{Chefles+1:98,Peres+1:98,Clarke+3:01,Bergou+2:04} serves this purpose. 

In the present context of ``asking particles where they have been''
(paraphrasing the title of Ref.~\cite{Danan+3:13}), 
unambiguous path knowledge is, as we will see, the most useful and, therefore,
Bob chooses the second strategy. 
After performing the unambiguous discrimination of the path-marker states 
$A\rho A\adj$,  $B\rho B\adj$, and $C\rho C\adj$, he places a bet on one of
the three cases whenever he is sure which checkpoint was visited.
If Bob wins all his bets, the path marker works correctly and the stored
which-path information is correctly extracted.

For the sake of simplicity and clarity of presentation, we assume that the
initial path-marker state is pure and describe it by a wave function $\psi$,
${\rho\repr\psi\psi\adj}$.
The action of $A$, $B$, or $C$ yields corresponding wave functions $\wf{A}$,
$\wf{B}$, or $\wf{C}$, each slightly different from $\psi$, in accordance with
\begin{equation}\label{eq:4-2}
  A\rho A\adj\repr\wf{A}\wf[\adj]{A}\,,\quad
  B\rho B\adj\repr\wf{B}\wf[\adj]{B}\,,\quad
  C\rho C\adj\repr\wf{C}\wf[\adj]{C}\,.
\end{equation}
Following Vaidman \cite{Vaidman:13a}, the ``pointer variables'' at checkpoints
A, B, and C are in the same initial state, with which we associate the wave
function $\chi_0^{\ }$, and the unitary operators $A$, $B$, and $C$ have the
same effect on their respective pointers, changing $\chi_0^{\ }$ to
$\chi_{\epsilon}^{\ }$.
Accordingly, we have the wave function 
$\psi=\chi_0^{\ }\otimes\chi_0^{\ }\otimes\chi_0^{\ }$ for the initial
path-marker state, and the final wave functions are
\begin{eqnarray}\label{eq:4-5a}
  \wf{A}&=&\chi_{\epsilon}^{\ }\otimes\chi_0^{\ }\otimes\chi_0^{\ }\,,
\nonumber\\
  \wf{B}&=&\chi_0^{\ }\otimes\chi_{\epsilon}^{\ }\otimes\chi_0^{\ }\,,
\nonumber\\
  \wf{C}&=&\chi_0^{\ }\otimes\chi_0^{\ }\otimes\chi_{\epsilon}^{\ }\,.
\end{eqnarray}
The expectation values of \Eq{3-12},
\begin{equation}\label{eq:4-5b}
  \left.\begin{array}{c}
  \expect{A}=\tr{\wf{A}\psi\adj}\\[1ex]  \expect{B}=\tr{\wf{B}\psi\adj}
  \end{array}\right\}
  =\chi_0\adj\chi_{\epsilon}\bigl(\chi_0\adj\chi_0^{\ }\bigr)^2
  =\chi_0\adj\chi_{\epsilon}\,,
\end{equation}
establish $\chi_0\adj\chi_{\epsilon}=\sqrt{1-3\epsilon}$, so that
\begin{equation}\label{eq:4-4}
  \wf[\adj]{a}\wf{b}=1-3\epsilon+3\epsilon\delta^{\ }_{\mathrm{ab}}
\quad\mbox{for a,b=A,B,C},
\end{equation}
which confirms that checkpoints A, B, and C are on equal footing.
We note that $(\wf{B}-\wf{A})\adj\wf{C}=0$ is a consequence of the symmetric
way of treating the three checkpoints.

Since only the three wave functions $\wf{A}$, $\wf{B}$, and $\wf{C}$ are
involved, we can use three-component columns for them.
A specific choice is
\begin{equation}\label{eq:4-3}
  \left.\begin{array}{c}
      \wf{A} \\[1ex] \wf{B}
    \end{array}\right\}=\sqcol{\pm\sqrt{3\epsilon/2}\\[0.5ex]%
                               -\sqrt{\epsilon/2}\\[0.5ex]%
                               \sqrt{1-2\epsilon}}\,,
  \quad \wf{C}=\sqcol{ 0 \\[0.5ex] \sqrt{2\epsilon} \\[0.5ex]%
                       \sqrt{1-2\epsilon}}\,,
\end{equation}
where we use square parentheses for the columns of the path-marker wave
functions to avoid confusion with the columns for the particle's path
amplitudes, such as the columns in \Eq{4-1}.

The error-minimizing measurement would allow Bob to guess right for a
fraction  
${\frac{1}{3}\bigl(\sqrt{1-2\epsilon}+2\sqrt{\epsilon}\bigr)^2}$ of the
particles \cite{Englert+1:10} while never being certain about the path.
But he wants to know the path for sure before placing his bet and, therefore, 
he employs the measurement for unambiguous discrimination with the
outcome operators $\Pi_{\mathrm{a}}=\mud{a}\mud[\adj]{a}$ for the three cases
a=A, B, or C and $\Pi_0=\phi_0^{\ }\phi_0\adj$ for the inconclusive outcome,
where \cite{Peres+1:98,Englert+1:10}
\begin{equation}\label{eq:4-6}
   \left.\begin{array}{c}
      \mud{A} \\[1ex] \mud{B}
    \end{array}\right\}=\sqcol{\pm\sqrt{1/2}\\[0.5ex]%
                               -\sqrt{1/6}\\[0.5ex]%
                               \sqrt{\epsilon/(3-6\epsilon)}}\,,
  \quad \mud{C}=\sqcol{ 0 \\[0.5ex] \sqrt{2/3} \\[0.5ex]%
                       \sqrt{\epsilon/(3-6\epsilon)}}
\end{equation}
and
\begin{equation}\label{eq:4-7}
  \phi^{\ }_0=\sqrt{\frac{1-3\epsilon}{1-2\epsilon}}\sqcol{0\\0\\1}\,.
\end{equation}
We have
\begin{eqnarray}\label{eq:4-8}
  \tr{\Pi^{\ }_{\mathrm{a}}\wf{b}\wf[\adj]{b}}
  &=&\bigl|\mud[\adj]{a}\wf{b}\bigr|^2=3\epsilon\delta^{\ }_{\mathrm{ab}}
\quad\mbox{for a,b=A,B,C}\,,\nonumber\\[1ex]
  \tr{\Pi^{\ }_0\wf{a}\wf[\adj]{a}}&=&1-3\epsilon\quad\mbox{for a=A,B,C}\,,
\end{eqnarray}
which tell us that Bob gets the inconclusive result for a fraction of
${1-3\epsilon}$ of the particles, and then he has no clue about the path, 
and does not place a bet.
For the remaining fraction of $3\epsilon$, he knows the path for sure and wins
his bet \cite{NoClue}.
As expected, the weak path marking (${\epsilon\ll1}$) does not store much
which-path information in the final path-marker state.

Here, then, is a more detailed account of the betting game alluded to above.
The source emits a particle into the interferometer and the unitary operators
$A$, $B$, and $C$, acting at checkpoints A, B, and C, respectively,
entangle the path qutrit of the particle with the marker degrees of freedom,
on which Bob performs the measurement of unambiguous discrimination.
He either gets the inconclusive result, in which case he does nothing, or he
finds one of the three conclusive results and then bets on the respective path
(``I bet that the particle went through checkpoint B,'' say).

Whenever Alice verifies the path, be it by the setup of \Fig{p-mode.a} or 
otherwise, she will confirm that Bob identified the actual path correctly.
This establishes, in a definite operational sense, that Bob's conclusive
results are in a strict one-to-one correlation with the particle's path
through the interferometer.
In this procedure, it does not matter whether Alice first detects the particle
and Bob later performs the unambiguous discrimination, or they do it in the
reverse temporal order. 

With beam splitters \BS{3} and \BS{4} in place, the final state of the
path marker, conditioned on detecting the particle by D, is that of \Eq{3-6},
\begin{equation}\label{eq:4-9}
  \final=\frac{1}{1+6\epsilon}\bigl(\wf{C}+\wf{B}-\wf{A}\bigr)
                              \bigl(\wf{C}+\wf{B}-\wf{A}\bigr)\adj\,.
\end{equation}
The conclusive outcomes of Bob's measurement for unambiguous discrimination
occur with equal probability,
\begin{equation}\label{eq:4-10}
  \tr{\Pi^{\ }_{\mathrm{a}}\final}=\frac{3\epsilon}{1+6\epsilon}
  \quad\mbox{for a=A,B,C},
\end{equation}
and
\begin{equation}\label{eq:4-11}
   \tr{\Pi^{\ }_0\final}=\frac{1-3\epsilon}{1+6\epsilon}
\end{equation}
is the conditional probability of getting the inconclusive outcome.
The equal probabilities of \Eq{4-10} are reminiscent of the equal
heights of the three peaks in \Fig{ExpData}.

Do these equal probabilities tell us that one-third of the particles went
through checkpoint A on the way from the source to the detector,
another one-third through checkpoint B, and the remaining one-third
through checkpoint C? 
No. This statement is correct only for those particles for which Bob obtains
definite path knowledge, which is the very small fraction 
$\frac{9\epsilon}{1+6\epsilon}$ of all particles.
Bob cannot say anything about the vast majority of the particles, for which he
gets the inconclusive outcome.

It is not possible to invoke a fair-sampling assumption here
(which is, however, an implicit assumption in Ref.~\cite{Danan+3:13}; see also
\cite{Vaidman:16a}) and so infer that the particles with unknown path are
as equally distributed over the three paths as those with a known path.
But it is possible to argue that Bob gets the inconclusive result for
particles that went through checkpoint C.


\subsection{An accounting exercise}\label{sec:WPK1b}
%
This argument has two ingredients.
First, we observe that an interferometer phase $\varphi$ introduced at 
checkpoint C, i.e., replace $\wf{C}$ by $\Exp{\I\varphi}\wf{C}$ in \Eq{4-9},
does not change the probability that the next particle is detected by D,
\begin{eqnarray}\label{eq:4-12}
   &&\expect{T_{\mathrm{fin}}\adj T_{\mathrm{fin}}^{\ }}
\nonumber\\&=&
  \frac{1}{9}\tr{\bigl(\Exp{\I\varphi}\wf{C}+\wf{B}-\wf{A}\bigr)
                \bigl(\Exp{\I\varphi}\wf{C}+\wf{B}-\wf{A}\bigr)\adj}
\nonumber\\
&=&\frac{1}{9}\wf[\adj]{C}\wf{C}
   +\frac{1}{9}(\wf{B}-\wf{A})\adj(\wf{B}-\wf{A})
   =\frac{1+6\epsilon}{9}\,,\qquad
\end{eqnarray}
where $(\wf{B}-\wf{A})\adj\wf{C}=0$ enters.
It follows that the amplitudes that reach beam splitter \BS{4} from
checkpoint C and \BS{3} are incoherent.
We must, therefore, add the probabilities rather than the probability
amplitudes. 

\begin{figure}
  \centering
  \includegraphics{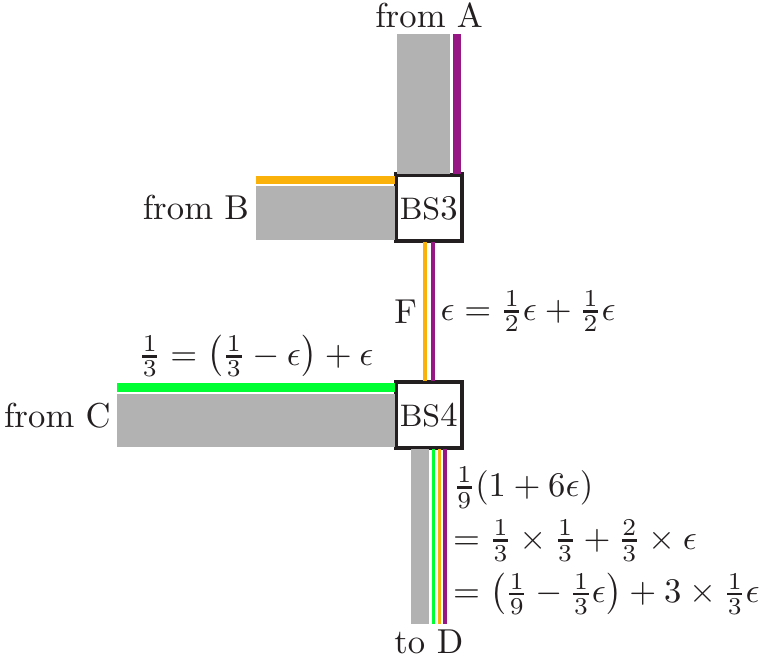}
  \caption{\label{fig:accounting}
      Unambiguous path knowledge about particles detected by detector D. 
      The next particle to enter the three-path interferometer is detected 
      with probability $\frac{1}{9}(1+6\epsilon)$; see \Eq{4-12}.
      On the way to detector D, the particle either passes checkpoint 
      C with probability
      $\frac{1}{3}$ and then has a $\frac{1}{3}$ chance of being reflected 
      by beam splitter \BS{4}, or it passes F with probability $\epsilon$
      and then has a $\frac{2}{3}$ chance of being transmitted by \BS{4}.
      The conclusive outcomes (purple, orange, green) of the unambiguous
      path discrimination  
      \emph{for these particles detected by D} identify the paths via
      checkpoints A, B, or C with equal probability of
      $\frac{1}{3}\epsilon\big/\frac{1}{9}(1+6\epsilon)$; see \Eq{4-10}. 
      Since that fully accounts for the particles that took the path
      \BS{3}$\to$\BS{4}$\to$D, the inconclusive measurement outcomes (gray)
      surely identify particles that followed the path C$\to$\BS{4}$\to$D.
      In the limit $\epsilon\to0$, \emph{all} particles reach D via C.
      The intensities at the other exit ports --- exit~\Path{i} from \BS{3}
      and exit~\Path{ii} from \BS{4} --- are not indicated in the figure.} 
\end{figure}

Second, the probabilities of reaching beam splitter \BS{4}
are $\frac{1}{3}$ (for path C$\to$\BS{4})
and $\epsilon$ (for path \BS{3}$\to$\BS{4}), and the
reflection and transmission probabilities of \BS{4} are
$\frac{1}{3}$ and $\frac{2}{3}$, respectively.
Indeed, the resulting probability
\begin{equation}\label{eq:4-13}
  \frac{1}{3}\times\frac{1}{3}+\frac{2}{3}\times\epsilon=\frac{1+6\epsilon}{9}
\end{equation}
agrees with that in \Eq{4-12}, with $\frac{1}{9}$ associated with checkpoint C
and $\frac{6\epsilon}{9}$ with the internal loop.
The corresponding fractions are
\begin{equation}\label{eq:4-14}
  \frac{1}{1+6\epsilon}
=\frac{1-3\epsilon}{1+6\epsilon}+\frac{3\epsilon}{1+6\epsilon}
\quad\mbox{for path C$\to$\BS{4}}
\end{equation}
and
\begin{equation}\label{eq:4-15}
  \frac{6\epsilon}{1+6\epsilon}=\frac{3\epsilon}{1+6\epsilon}
                               +\frac{3\epsilon}{1+6\epsilon}
\quad\mbox{for path \BS{3}$\to$\BS{4}.}
\end{equation}
The fraction for path \BS{3}$\to$\BS{4} is fully accounted for by
the conclusive outcomes for checkpoints A and B in \Eq{4-10}, and the
fraction for path C$\to$\BS{4} is the sum of the fraction for the
conclusive outcome C and the fraction for the inconclusive outcome 
in \Eq{4-11}.
This accounting exercise suggests strongly that the fraction 
$\frac{1-3\epsilon}{1+6\epsilon}$ of Bob's inconclusive outcomes is associated
with particles that passed checkpoint C.
If this were indeed so, all particles detected by D would have gone by  
checkpoint C in the limit of ${\epsilon\to0}$, fully consistent with the
common-sense reading of \Fig{fwd-bwd}.
The matter is summarized in Fig.~\ref{fig:accounting}.

We emphasize the great benefit of the unambiguous path discrimination 
by the measurement with the outcomes $\Pi_{\mathrm{A}}$,  $\Pi_{\mathrm{B}}$,
$\Pi_{\mathrm{C}}$, and $\Pi_0$ of Eqs.~(\ref{eq:4-6}) and (\ref{eq:4-7}): 
While the inconclusive outcome $\Pi_0$ yields no path information 
for the whole ensemble of particles emitted by the source S, it provides
definite path knowledge (via checkpoint C, that is) for the subensemble of
particles detected by detector D; more about this in the next section.
All of the pre-selected and post-selected particles of interest have a known
path through Vaidman's interferometer. 
We ask these particles where they have been, and they all give a definite
answer.

Whether one finds this reasoning convincing or regards it as another
appeal to common sense of no consequence, it is certainly the case
that the outcomes $\Pi_{\mathrm{A}}^{\ }$ and $\Pi_{\mathrm{B}}^{\ }$ account fully 
for the fraction of particles that reach beam splitter \BS{4} via 
checkpoints A or B.  
How do we reconcile the definitely known paths with the destructive
interference at \BS{3}?
True, destructive interference is only possible if the path cannot be known
and, yet, there is no contradiction here, as we shall see in
Sec.~\ref{sec:2paths}.

In passing, we note another consequence of the lack of coherence between the
amplitudes for paths \BS{3}$\to$\BS{4} and C$\to$\BS{4}.
If we block the path from checkpoint C to beam splitter \BS{4} (or,
equivalently, remove the entrance beam splitter \BS{1}), then we get a signal
with a strength proportional to $\epsilon$ at detector D.
This is in marked contrast to the experiment of Danan \textit{et al.\/}
\cite{Danan+3:13}, where the corresponding amplitudes are coherent --- the
electric field vectors of the arriving partial beams just add to yield the
fields of the emerging beams --- and, therefore, the power spectrum in their
Fig.~2(b) is proportional to their analog of $\epsilon$ while that in
Fig.~2(c) is proportional to $\epsilon^2$ and buried in the noise; 
Fig.~4 in \cite{Sokolovski:17a} illustrates this point.

\subsection{Sorted subensembles}\label{sec:WPK1c}
At an instant after unitary operators $A$, $B$, $C$ acted when the particle
passed checkpoints A, B, C and before the path amplitudes are processed by beam
splitters \BS{3} and \BS{4}, the entangled state of the path marker and the
particle has the wave function 
\begin{eqnarray}\label{eq:4-16}\label{eq:8-1}
  &&\frac{1}{\sqrt{3}}{\left[\wf{A}\otimes\column{1\\0\\0}
 +\wf{B}\otimes\column{0\\1\\0}
 +\wf{C}\otimes\column{0\\0\\1}\right]}\nonumber\\
&=&\frac{1}{\sqrt{3}}\column{\wf{A} \\ \wf{B} \\ \wf{C}}\,.
\end{eqnarray}
Upon performing his measurement for unambiguous path discrimination, Bob sorts
the particles into subensembles in accordance with the outcome he observes
\cite{sorting}. 
The density matrices for the subensembles with known paths (purple, orange,
green in \Fig{accounting}),
\begin{equation}\label{eq:4-17}
  \frac{1}{3}\column{\mud[\adj]{a}\wf{A}\\ %
                     \mud[\adj]{a}\wf{B}\\ %
                     \mud[\adj]{a}\wf{C}}
             \column{\mud[\adj]{a}\wf{A}\\ %
                     \mud[\adj]{a}\wf{B}\\ %
                     \mud[\adj]{a}\wf{C}}\adj
  =\frac{1}{3}\mud[\adj]{a}\column{\wf{A}\\ \wf{B}\\ \wf{C}}
              \column{\wf{A}\\ \wf{B}\\ \wf{C}}\adj\mud{a}
\end{equation}
for a=A,B,C, are rank-one projectors on their respective paths with 
\begin{eqnarray}\label{eq:4-18}
  \frac{1}{\sqrt{3}}\mud[\adj]{A}\column{\wf{A}\\ \wf{B}\\ \wf{C}}
  &=&\sqrt{\epsilon}\column{1\\ 0\\ 0}\,,\nonumber\\
  \frac{1}{\sqrt{3}}\mud[\adj]{B}\column{\wf{A}\\ \wf{B}\\ \wf{C}}
  &=&\sqrt{\epsilon}\column{0\\ 1\\ 0}\,,\nonumber\\
  \frac{1}{\sqrt{3}}\mud[\adj]{C}\column{\wf{A}\\ \wf{B}\\ \wf{C}}
  &=&\sqrt{\epsilon}\column{0\\ 0\\ 1}\,;
\end{eqnarray}
the strict one-to-one correspondence between Bob's outcome and the particle's
path is manifest here. 
We keep the probability amplitudes of $\sqrt{\epsilon}$ as overall
factors in the particle wave functions to remind us of the statistical weights
of the subensembles, as indicated by the thickness of the purple, orange, and
green lines in \Fig{accounting}. 

These three subensembles of particles with operationally known paths are
supplemented by the fourth subensemble for Bob's inconclusive outcome (gray in
\Fig{accounting}). 
Its density matrix is also a rank-one projector with the wave function
\begin{equation}\label{eq:4-19}
 \frac{1}{\sqrt{3}}\phi\adj_0\column{\wf{A}\\ \wf{B}\\ \wf{C}}
 =\sqrt{\frac{1-3\epsilon}{3}}\column{1 \\ 1 \\ 1}\,, 
\end{equation}
which is exactly the ${\epsilon=0}$ wave function of particles that are not
monitored while traversing the three-path interferometer [cf.\ \Eq{A-6}], 
multiplied by the probability amplitude $\sqrt{1-3\epsilon}$ for the
inconclusive outcome. 
No path knowledge is available for the particles in this subensemble.

The wave functions in \Eq[Eqs.~]{4-16}, \Eq[]{4-18}, and \Eq[]{4-19} apply
before the particle reaches beam splitter \BS{3}.
The wave functions after \BS{3} and before \BS{4} are
\begin{equation}\label{eq:4-20a}
  U_3\frac{1}{\sqrt{3}}\column{\wf{A} \\ \wf{B} \\ \wf{C}}=
  \frac{1}{\sqrt{6}}
  \column{\wf{B}+\wf{A} \\ \wf{B}-\wf{A} \\ \sqrt{2}\,\wf{C}}
\end{equation}
for the entangled marker-particle state, and
\begin{equation}\label{eq:4-20b}
  \frac{1}{\sqrt{6}}\mud[\adj]{x}
  \column{\wf{B}+\wf{A} \\ \wf{B}-\wf{A} \\ \sqrt{2}\,\wf{C}}
\quad\text{for x=A,B,C, or 0}
\end{equation}
for the conditional particle states.
In view of the eventual conditioning on particles detected by detector D, we
focus on the components that are processed by beam splitter \BS{4} and
project out the then irrelevant path-\Path{i} component,
\begin{equation}\label{eq:4-21}
  \column[ccc]{0&0&0\\0&1&0\\0&0&1}
  \frac{1}{\sqrt{6}}\mud[\adj]{x}
  \column{\wf{B}+\wf{A} \\ \wf{B}-\wf{A} \\ \sqrt{2}\,\wf{C}}
 =\column{0\\ \frac{1}{\sqrt{6}}\mud[\adj]{x}(\wf{B}-\wf{A}) \\ 
              \frac{1}{\sqrt{3}}\mud[\adj]{x}\wf{C}}
\end{equation}
and so arrive at the conditional wave functions for the four subensembles,
namely
\begin{equation}\label{eq:4-22}
   \sqrt{\frac{\epsilon}{2}}\column{0\\-1\\0}\,,\quad
   \sqrt{\frac{\epsilon}{2}}\column{0\\1\\0}\,,\quad
   \sqrt{\epsilon}\column{0\\0\\1}
\end{equation}
for x=A, B, and C, respectively, and    
\begin{equation}\label{eq:4-23}
\sqrt{\frac{1-3\epsilon}{3}}\column{0\\0\\1}
\end{equation}
for the inconclusive outcome.
This confirms the conclusion reached by the accounting exercise of
Sec.~\ref{sec:WPK1b}: 
Whenever Bob obtains the inconclusive outcome for a particle detected by
detector D, then this particle has arrived at beam splitter \BS{4} along
path~\Path{iii}, i.e., via checkpoint C \cite{fn:EOM3}.

Finally, for the record, the entangled marker-particle state at the exit stage
of the three-path interferometer --- before the particle is detected at exit
\Path{i}, \Path{ii}, or \Path{iii} --- is
\begin{equation}\label{eq:4-24a}
  U_4U_3\frac{1}{\sqrt{3}}\column{\wf{A} \\ \wf{B} \\ \wf{C}}=
  \column{\frac{1}{\sqrt{6}}(\wf{B}+\wf{A}) \\
          \frac{1}{\sqrt{18}}(2\wf{C}-\wf{B}+\wf{A}) \\
          \frac{1}{3}(\wf{C}+\wf{B}-\wf{A})}\,,
\end{equation} 
and the particle wave functions associated with the paths through checkpoints
A, B, and C are
\begin{equation}\label{eq:4-24b}
  \sqrt{\frac{\epsilon}{6}}\column{\sqrt{3} \\ 1 \\ -\sqrt{2}}\,,\quad
  \sqrt{\frac{\epsilon}{6}}\column{\sqrt{3} \\ -1 \\ \sqrt{2}}\,,\quad
  \sqrt{\frac{\epsilon}{3}}\column{0 \\ \sqrt{2} \\ 1}\,,
\end{equation}
respectively, while the wave function
\begin{equation}\label{eq:4-24c}
  \frac{1}{3}\sqrt{1-3\epsilon}\column{\sqrt{6} \\ \sqrt{2} \\ 1}
\end{equation}
applies to the particles with unknown path.
Alice can check which one is the actual path by an interferometric measurement
that discriminates between the three orthogonal wave functions in \Eq{4-24b}.
This is an example of the ``otherwise'' in the betting game between
\Eq[Eqs.~]{4-8} and \Eq[]{4-9}.

\begin{figure}
  \centering
  \includegraphics{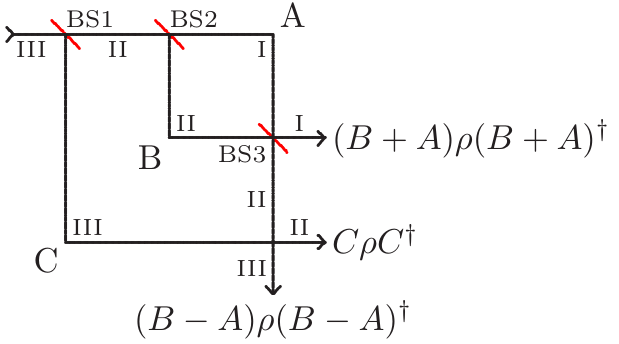}
  \caption{\label{fig:p-mode.b}%
    Marking the path through the interferometer as in \Fig{p-mode.a}.
    Now, beam splitter \BS{3} is in place but \BS{4} is still removed.
    The conditional states of the marker degrees of freedom are
    $(B+A)\rho(B+A)\adj$, $C\rho C\adj$, and $(B-A)\rho(B-A)\adj$, 
    respectively, upon detecting the particle at exit \Path{i}, \Path{ii},
    or \Path{iii}.}
\end{figure}

\section{More which-path knowledge}\label{sec:WPK2}
Bob's measurement with the outcome operators specified by \Eq[Eqs.~]{4-6} and
\Eq[]{4-7} distinguishes between the three possible detection events of the
setup in \Fig{p-mode.a}. 
In view of \Eq{4-12}, it appears to be more useful to tell apart the three
detection events of the setup in \Fig{p-mode.b}, where beam splitter
\BS{3} is in place while \BS{4} is not.
Alice and Bob then play the betting game described between \Eq[Eqs.~]{4-8} and
\Eq[]{4-9}, with the necessary modifications as they follow from
\begin{equation}\label{eq:5-1}
   \frac{1}{6}\column{B+A \\ \sqrt{2}\,C \\ B-A}\rho
              \column{B+A \\ \sqrt{2}\,C \\ B-A}\adj
\end{equation}
for the joint state of the particle and the path marker instead of the state
in \Eq{4-1}.

Here, Bob's task is to discriminate between the three path-marker states
\begin{eqnarray}\label{eq:5-2}
  \left.\begin{array}{c}\rho_{\sPath{i}} \\
                        \rho_{\sPath{iii}}\end{array}\right\}
   &\propto& (B\pm A)\rho(B\pm A)\adj=
             (\wf{B}\pm\wf{A})(\wf{B}\pm\wf{A})\adj\,,\nonumber\\
   \rho_{\sPath{ii}}  &\propto& C\rho C\adj=\wf{C}\wf[\adj]{C}\,,
\end{eqnarray}
which have a priori probabilities of occurrence of
\begin{eqnarray}\label{eq:5-3}
  \frac{1}{6}\tr{(B\pm A)\rho(B\pm A)\adj}&=&
  \left\{\begin{array}{l}%
         \displaystyle\frac{1}{3}(2-3\epsilon)\,,\\[1.5ex]%
         \epsilon\,,\end{array}\right.
\nonumber\\ 
   \frac{2}{6}\tr{C\rho C\adj}&=&\frac{1}{3}\,,
\end{eqnarray}
respectively [cf.\ \Eq{4-20a}].
Much more useful than discriminating between $\rho_{\sPath{i}}$,
$\rho_{\sPath{ii}}$, and $\rho_{\sPath{iii}}$, however, is a measurement
that tells $\rho_{\sPath{ii}}$ and $\rho_{\sPath{iii}}$ apart because 
$\rho_{\sPath{i}}$ is irrelevant for the particles detected by detector D in
the full setup of \Fig{3paths} with beam splitter \BS{4} in place.
Accordingly, the appropriate verification protocol is this:
Alice takes note of the particles detected at exits \Path{ii} and
\Path{iii} in the setup of \Fig{p-mode.b}, and challenges Bob to place a bet.

Since $(\wf{B}-\wf{A})\adj\wf{C}=0$, Bob uses an orthogonal measurement where
the outcome operators $\Pi_j=\phi^{\ }_{j}\phi_{j}\adj$ for
$j=\Path{i,ii,iii}$ project on 
\begin{eqnarray}\label{eq:5-4}
  \phi^{\ }_{\sPath{i}}&=&\sqcol{0\\-\sqrt{1-2\epsilon}\\\sqrt{2\epsilon}}
  =\frac{\wf{A}+\wf{B}-2(1-3\epsilon)\wf{C}}%
        {3\sqrt{2\epsilon(1-2\epsilon)}}\,,
\nonumber\\
  \phi^{\ }_{\sPath{ii}}&=&\sqcol{0\\\sqrt{2\epsilon}\\\sqrt{1-2\epsilon}}
      =\wf{C}\,,
\nonumber\\
  \phi^{\ }_{\sPath{iii}}&=&\sqcol{1\\0\\0}
=\frac{\wf{A}-\wf{B}}{\sqrt{6\epsilon}}\,,
\end{eqnarray}
which are such that 
\begin{eqnarray}\label{eq:5-5}
\tr{\Pi_{\sPath{i}}\rho_{\sPath{ii}}}=0\,,
&\quad&\tr{\Pi_{\sPath{i}}\rho_{\sPath{iii}}}=0\,,\nonumber\\
\tr{\Pi_{\sPath{ii}}\rho_{\sPath{ii}}}=1\,,
&\quad&\tr{\Pi_{\sPath{ii}}\rho_{\sPath{iii}}}=0\,,\nonumber\\
\tr{\Pi_{\sPath{iii}}\rho_{\sPath{ii}}}=0\,,
&\quad&\tr{\Pi_{\sPath{iii}}\rho_{\sPath{iii}}}=1\,.
\end{eqnarray}
It follows that Bob will never get outcome~\Path{i}; he bets on 
exit~\Path{ii} when he gets outcome~\Path{ii}, and on 
exit~\Path{iii} for outcome~\Path{iii}. 
As there are no inconclusive measurement outcomes,
Bob will place a bet for every particle detected by Alice 
in exit~\Path{ii} or exit~\Path{iii}, and he wins all bets.
That is: He knows for sure, for each and every particle, whether it went
through checkpoint C (outcome~\Path{ii}) or took the route through the
internal loop (outcome~\Path{iii}).

Now, with beam splitter \BS{4} back in place, Bob examines the final
path-marker state of \Eq{4-9} and he gets outcomes \Path{i}, \Path{ii},
\Path{iii} with the probabilities 
\begin{eqnarray}\label{eq:5-6}
  \tr{\Pi_{\sPath{i}}\final}&=&0\,,\nonumber\\
  \tr{\Pi_{\sPath{ii}}\final}&=&\frac{1}{1+6\epsilon}\,,\nonumber\\
  \tr{\Pi_{\sPath{iii}}\final}&=&\frac{6\epsilon}{1+6\epsilon}\,.
\end{eqnarray}
As we know, the particle's probability amplitudes processed by beam splitter
\BS{4} refer to known paths --- this is the essence of \Eq{5-5} --- and,
therefore, they are incoherent and we add the probabilities rather than
the probability amplitudes.
As we did in \Eq{4-13}, we account for the reflection and transmission
probabilities of \BS{4}, establish the a priori probabilities
of being detected by D as $\frac{1}{9}$ and $\frac{2}{3}\epsilon$ for the
particle arriving via checkpoint C or the internal loop, respectively, and
so confirm that the probabilities in \Eq{5-6} are the correct relative
frequencies, 
\begin{equation}\label{eq:5-7}
  \frac{1}{1+6\epsilon}
  =\frac{\frac{1}{9}}{\frac{1}{9}+\frac{2}{3}\epsilon}\,,
\quad
  \frac{6\epsilon}{1+6\epsilon}
  =\frac{\frac{2}{3}\epsilon}{\frac{1}{9}+\frac{2}{3}\epsilon}\,.
\end{equation}
In the limit ${\epsilon\to0}$, Bob gets  outcome~\Path{ii} for every
particle, i.e., every particle arrives at detector D via checkpoint C. 
This is exactly what common sense tells us.

This confirms once more our conclusions reached by the accounting exercise of
Sec.~\ref{sec:WPK1b} and the sorted-subensembles argument of
Sec.~\ref{sec:WPK1c}:
In the measurement of \Eq[Eqs.~]{4-6} and \Eq[]{4-7}, all inconclusive results
are obtained for particles that passed checkpoint C before reaching 
detector D; this measurement, then, provides full path knowledge for every
particle detected by D.
The extrapolation ${\epsilon\to0}$ supports the common-sense conclusion in
Sec.~\ref{sec:3paths}, which simply recognizes that there is only one blue path
from source S to detector D in \Fig{fwd-bwd}.

With regard to the final path-marker wave function in \Eq{4-9}, these
considerations establish that we should read it as the superposition of the
normalized wave functions $\wf{C}$ for the path C$\to$\BS{4}$\to$D and
$(\wf{B}-\wf{A})/\sqrt{6\epsilon}$ for the path \BS{3}$\to$\BS{4}$\to$D,
that is
\begin{eqnarray}\label{eq:5-8}
  && \frac{1}{\sqrt{1+6\epsilon}}(\wf{C}+\wf{B}-\wf{A})\nonumber\\
  &=&\frac{1}{\sqrt{1+6\epsilon}}\wf{C}
   +\sqrt{\frac{6\epsilon}{1+6\epsilon}}\frac{\wf{B}-\wf{A}}{\sqrt{6\epsilon}}
  \,.
\end{eqnarray}
Since $\wf{C}=\phi^{\ }_{\sPath{ii}}$ and
$\wf{B}-\wf{A}=-\sqrt{6\epsilon}\,\phi^{\ }_{\sPath{iii}}$ are orthogonal, this
superposition refers to fully distinguishable alternatives [see \Eq{5-6}], and
in the limit of ${\epsilon\to0}$ there is only the path C$\to$\BS{4}$\to$D.

\begin{figure*}
  \centering
  \includegraphics{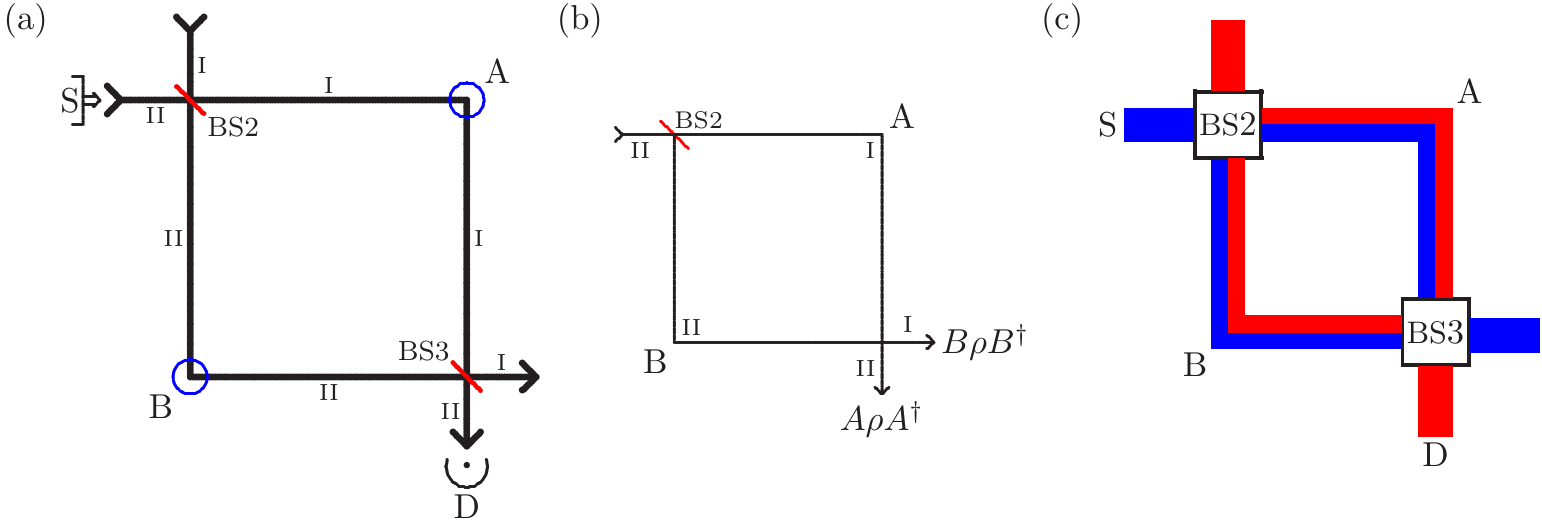}
  \caption{\label{fig:2paths}Two-path interferometer. 
    (a) Particles are emitted
    from source S, detected by detector D, and pass checkpoints 
    A or B on the way from S to D.  
    (b) Beam splitter \BS{3} is removed for the verification that the
    discrimination between the final path-marker states $A\rho A\adj$ and
    $B\rho B\adj$ correctly identifies the paths via checkpoints A and B. 
    (c) In this analog of \Fig{fwd-bwd}, the blue lines indicate the
    probabilities for the particles that surely emerged from source S,
    and the red lines are for the particles that will surely be detected by 
    detector D.}
\end{figure*}

\section{Two-path interferometer with weak path marking}
\label{sec:2paths}
We return to the question raised at the end of Sec.~\ref{sec:WPK1} and focus
on the two-path interferometer of the internal loop.
The scheme of a two-path experiment is sketched in \Fig{2paths}(a) where the
unitary matrix for beam splitters \BS{2} and \BS{3} is that of \Eq{2-1},
only that we ignore the third row and column in the present context, so that
\begin{equation}\label{eq:6-1}
  U_2=U_3=\frac{1}{\sqrt{2}}\column[cc]{1&1\\-1&1}
\end{equation}
here.
For the path marking, too, we now use a binary degree of freedom (``qubit'')
and represent the unitary operators $A$ and $B$ as well as the initial
path-marker state $\rho$ by two-by-two matrices. 
In particular, we choose
\begin{eqnarray}\label{eq:6-2}
  \left.\begin{array}{c}
    A\rho A\adj \\[1ex]  B\rho B\adj
  \end{array}\right\}
   &\repr&\sqcol{\sqrt{1-\epsilon} \\ \mp\sqrt{\epsilon}}
      \sqcol{\sqrt{1-\epsilon} \\ \mp\sqrt{\epsilon}}\adj\,,
\nonumber\\
  A\rho B\adj &\repr&\sqcol{\sqrt{1-\epsilon} \\ -\sqrt{\epsilon}}
      \sqcol{\sqrt{1-\epsilon} \\ \sqrt{\epsilon}}\adj\,,
\end{eqnarray}
where $\epsilon$ is the probability that the next particle emitted by 
source S will reach detector D,
\begin{equation}\label{eq:6-3}
  \epsilon=\tr{T_{\mathrm{fin}}^{\ }\rho T_{\mathrm{fin}}\adj}
\end{equation}
with $T_{\mathrm{fin}}^{\ }=\half(B-A)$.
After detecting the particle, the conditional final state of the path-marker
qubit is
\begin{equation}\label{eq:6-4}
  \final=\frac{1}{\epsilon}T_{\mathrm{fin}}^{\ }\rho T_{\mathrm{fin}}\adj
\repr\sqcol[cc]{0 & 0 \\ 0 & 1}\,.
\end{equation}

\Fig[Figure~]{2paths}(b) depicts the situation with beam splitter \BS{3}
removed, in which we implement the analog of the protocol for \Fig{p-mode.a}:
Alice records whether the particle took exit~\Path{i} or exit~\Path{ii}, and
Bob measures the path-marker qubit and then places a bet. 
His measurement for unambiguous discrimination has the outcome operators
\begin{equation}\label{eq:6-5}
   \left.\begin{array}{c}
    \Pi^{\ }_{\mathrm{A}} \\[1ex]   \Pi^{\ }_{\mathrm{B}}
  \end{array}\right\}\repr
 \half\sqcol[c@{\enskip}c]{\ds\frac{\epsilon}{1-\epsilon} 
                 & \ds\mp\sqrt{\frac{\epsilon}{1-\epsilon}} \\[2.5ex]
                   \ds\mp\sqrt{\frac{\epsilon}{1-\epsilon}} & 1}
\end{equation}
for the paths identified by checkpoints A and B, and
\begin{equation}\label{eq:6-6}
  \Pi^{\ }_0\repr\frac{1-2\epsilon}{1-\epsilon}\sqcol[cc]{1&0\\0&0}
\end{equation}
for the inconclusive measurement.
Bob places a bet for every conclusive outcome (fraction $2\epsilon$ of all
cases) and wins all the bets.
He does not bet on the exit when he gets the inconclusive outcome
(fraction ${1-2\epsilon}$). 

With beam splitter \BS{3} in place and the particle detected by D at
exit~\Path{ii}, Bob's outcome probabilities are
\begin{eqnarray}\label{eq:6-7}
  p^{\ }_{\mathrm{A}}&=&\tr{\Pi^{\ }_{\mathrm{A}}\final}=\half  \,,\nonumber\\
  p^{\ }_{\mathrm{B}}&=&\tr{\Pi^{\ }_{\mathrm{B}}\final}=\half  \,,\nonumber\\
  p^{\ }_0&=&\tr{\Pi^{\ }_0\final}=0\,.
\end{eqnarray}
Since there are no inconclusive outcomes, Bob knows for \emph{every}
particle detected by D whether it went through checkpoint A or
checkpoint B on its way from the source to the detector.

If, rather than detecting the particle at exit~\Path{ii}, we detect it at
exit~\Path{i}, which will happen with probability ${1-\epsilon}$ for the
next particle, the conditional final state of the path marker is
\begin{equation}\label{eq:6-8}
  \rho'_{\mathrm{fin}}=\frac{1}{1-\epsilon}\frac{B+A}{2}\rho\frac{(B+A)\adj}{2}
        \repr\sqcol[cc]{1&0\\0&0}\,,
\end{equation}
and Bob obtains his measurement outcomes with the probabilities
\begin{equation}\label{eq:6-9}
  p'_{\mathrm{A}}=p'_{\mathrm{B}}=\half\frac{\epsilon}{1-\epsilon}
  \quad\mbox{and}\quad
  p'_0=\frac{1-2\epsilon}{1-\epsilon}\,.
\end{equation}
That is: The small fraction $\frac{\epsilon}{1-\epsilon}\simeq\epsilon$ of the
particles that take exit~\Path{i} have a known path through the
interferometer while the path is unknown for the vast majority of the
particles (fraction $\frac{1-2\epsilon}{1-\epsilon}\simeq1-\epsilon$). 
These particles with unknown, and unknowable, paths exhibit interference ---
perfect constructive interference for exit~\Path{i}, perfect destructive
interference for exit~\Path{ii}. 

Accordingly, we have this full picture for the particles observed after
exiting from the interferometer:
A fraction ${p^{\ }_{\mathrm{A}}\epsilon+p'_{\mathrm{A}}(1-\epsilon)=\epsilon}$
of all particles passed checkpoint A and half of them emerged from
exit~\Path{i}, the other half from exit~\Path{ii}.
Another fraction 
${p^{\ }_{\mathrm{B}}\epsilon+p'_{\mathrm{B}}(1-\epsilon)=\epsilon}$ surely passed
checkpoint B and half of them took exit~\Path{i}, the other half
exit~\Path{ii}. 
The remaining fraction
$p^{\ }_0\epsilon+p'_0(1-\epsilon)= 1-2\epsilon$ consists entirely of
particles with unknowable paths through the interferometer and full
interference strength; they all emerged from exit~\Path{i}.

We can change the distribution of the interfering particles between the exits
by introducing a phase $\varphi$ into the interferometer --- formally by the
replacement $A\rho B\adj\to\Exp{\I\varphi}A\rho B\adj$ in \Eq{6-2}.
Then we observe interference fringes with a visibility of ${1-2\epsilon}$. 

Here, then, is the answer to the question asked in the next-to-last
paragraph of Sec.~\ref{sec:WPK1b}:
The particles that pass checkpoint F in the three-path interferometer of
\Fig{3paths} do not participate in the interference; all the interfering
particles take exit~\Path{i} at beam splitter \BS{3}.
There really is no contradiction.

We arrive at this full picture thanks to the which-path information
acquired by the unambiguous path discrimination, which enables us to sort
the  particles into subensembles with either a surely-known path or an
unknowable path.   
The error-minimizing measurement would provide information of another kind
that is useful for its own purpose but does not yield a definite answer when
``asking particles where they have been.''

As a final remark we note that the analog of \Fig{fwd-bwd} in \Fig{2paths}(c)
shows coexisting blue and red lines for the interferometer loop but not
between source S and beam splitter \BS{2} (no red probability) nor
between beam splitter \BS{3} and detector D (no blue probability). 
Shall we, therefore, adopt the narrative that the particle was inside the loop
but did not enter or leave? Certainly not.
Further, the weak values of the projectors on paths~\Path{i}
and~\Path{ii} at checkpoints A and B are ill-defined unless there is an
interferometer phase with $\Exp{\I\varphi}\neq1$.
Then these weak values are ${\half\pm\frac{1}{2\I}\cot\frac{\varphi}{2}}$, and
\Fig{2paths}(c) applies for ${0\neq\varphi\to0}$.
We do not offer a physical interpretation of these weak values and leave the
matter at that \cite{weak2}.

\begin{figure}
  \centering
  \includegraphics{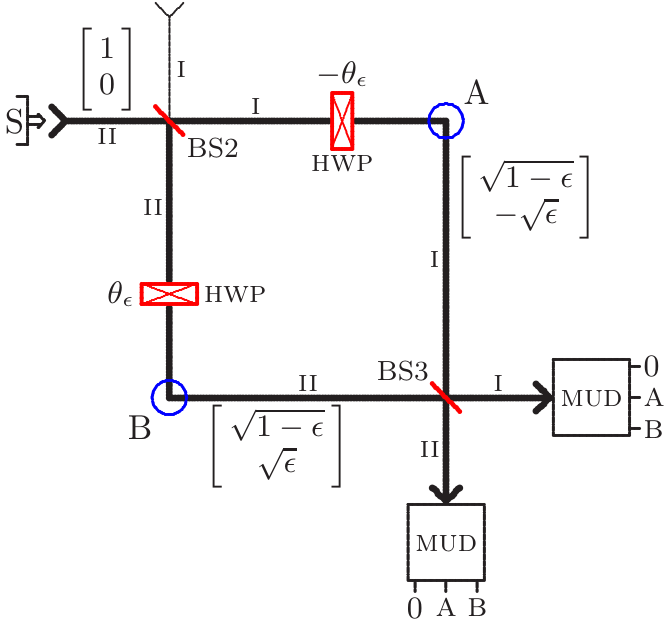}
  \caption{\label{fig:same-photon}%
    Single-photon two-path interferometer with the path marked on the
    polarization. 
    Source S emits one photon at a time, vertically polarized, into the
    interferometer. 
    The half-wave plates (\textsc{hwp}s) rotate the polarization slightly,
    such that the amplitude for vertical polarization is $\sqrt{1-\epsilon}$
    in both paths, while the horizontal polarization has amplitude
    $\sqrt{\epsilon}$ for the photons passing through checkpoint A, and
    amplitude $-\sqrt{\epsilon}$ for those passing through checkpoint B.
    At each exit port, a measurement for unambiguous discrimination 
    (\textsc{mud}, see \Fig{MUD}) between the two polarization states is
    executed when the photon is detected.}
\end{figure}

\section{Single-photon two-path interferometer %
            with marked paths  and unambiguous path knowledge}
\label{sec:proposal}
\subsection{The interfering photon carries the polarization}\label{sec:same}
One experimental realization of the two-path interferometer of \Fig{2paths}
with photons as the interfering particles is sketched in \Fig{same-photon}, 
where we use the photon's polarization degree of freedom as the path-marker
qubit. 
The path-marker amplitudes $v$ and $h$ in 
\mbox{\small$\Bigl[\begin{array}{c}v\\[-1ex]h\end{array}\Bigr]$}
now refer to the vertical and horizontal polarization, respectively, and all
photons are vertically polarized before entering the interferometer loop at
beam splitter \BS{2}.

Inside the interferometer, we have half-wave plates (\textsc{hwp}s) set at
angles $-\theta_{\epsilon}$ in path~\Path{i} and $\theta_{\epsilon}$ in
path~\Path{ii}, with $\cos(2\theta_{\epsilon})=\sqrt{1-\epsilon}$ 
and $\sin(2\theta_{\epsilon})=\sqrt{\epsilon}$.
The effect of a half-wave plate on the two-component column of polarization
amplitudes is described by the two-by-two matrix in
\begin{equation}\label{eq:7-1}
  \sqcol{v\\h}\mathrel{\begin{array}{@{}l@{}}%
                       \mbox{\small\ \textsc{hwp}}\\[-1.5ex]%
                       \mbox{\LARGE$\longrightarrow$}\\[-2.5ex]%
                       \mbox{\footnotesize{}\ at $\theta$}\end{array}}
  \sqcol[cc]{\cos(2\theta) & \sin(2\theta) \\ 
             \sin(2\theta) & -\cos(2\theta)}
   \sqcol{v\\h}\,.
\end{equation}
Therefore, we have the columns of \Eq{6-2} for the polarizations of the
photons after the half-wave plates, so that these polarizations are associated
with the respective checkpoints.
At each exit, the photons are detected by a measurement for unambiguous
discrimination (\textsc{mud}) that has the three outcome operators of
\Eq[Eqs.~]{6-5} and \Eq[]{6-6}; 
its realization is described below (see \Fig{MUD}).

In the verification mode of \Fig{2paths}(b), beam splitter \BS{3} is removed. 
Alice knows which measurement for unambiguous discrimination detected the
photon, while Bob is told which outcome was found ($0$ or A or B) but not at
which exit.

The single-photon experiment of \Fig{same-photon} correctly implements the
scheme of \Fig{2paths} but it suffers from the drawback that the interfering
particle is also the carrier of the path-marker qubit.
We cannot detect the particle ($\equiv$ the photon) without at the same time
measuring the path-marker qubit ($\equiv$ its polarization).
Then, the distinction between observers Alice and Bob is rather artificial: 
The  verification described in the preceding paragraph would require an
arbiter who, after noting at which of the six outcomes of both measurements
for unambiguous discrimination the photon was registered, 
informs Alice about the exit and Bob about the outcome.
 
Therefore, an experimental implementation in which the path qubit and the
path-marker qubit are physically separated and can be measured individually is
desirable.
We describe such an implementation in the next section.

\begin{figure}
  \centering
  \includegraphics{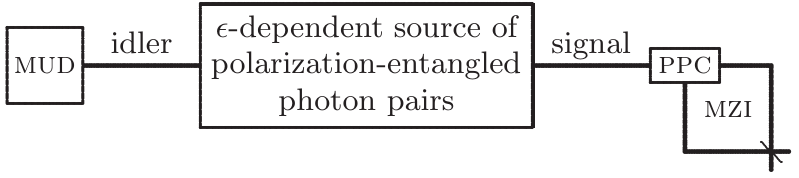}
  \caption{\label{fig:photonpair}%
    Scheme of the photon-pair experiment.
    The source emits an idler-signal pair of polarization-entangled photons in a
    state determined by the $\epsilon$ parameter of \Eq{6-2}.   
    The signal enters the Mach--Zehnder interferometer (\textsc{mzi}) 
    with a polarization-to-path converter (\textsc{ppc}, see \Fig{convert}) at
    the entrance. 
    The idler polarization is correlated with the signal path through
    the \textsc{mzi}.  
    The measurement for unambiguous discrimination (\textsc{mud}, see \Fig{MUD})
    extracts this path information from the idler polarization.
    }
\end{figure}

\subsection{A partner photon carries the polarization}
\label{sec:signal-idler}
The scheme is presented in \Fig{photonpair}.
The two-photon source is of the kind pioneered by Kwiat \textit{et al.\/}
\cite{Kwiat+4:99} (see also \cite{Dai+4:13}), in which spontaneous parametric
down-conversion produces photon pairs (``signal'' and ``idler'') in a
polarization-entangled state with the wave function
\begin{equation}\label{eq:7-2}
    \sqrt{1-\epsilon}\sqcol{1\\0}\otimes\sqcol{1\\0}
    +\sqrt{\epsilon}\sqcol{0\\1}\otimes\sqcol{0\\1}\,,
\end{equation}
where the first factor in the tensor product is for the idler polarization and
the second for the signal polarization.
Consistent with the meaning of the two-component columns in \Eq{7-1} and
\Fig{same-photon}, the probability for finding both photons vertically
polarized is ${1-\epsilon}$, and that for both horizontally polarized is
$\epsilon$. 
Before leaving the source, the signal traverses a half-wave plate set
at $22.5^{\circ}$, which turns the wave function into
\begin{eqnarray}\label{eq:7-3}
   &&\sqrt{\frac{1-\epsilon}{2}}\sqcol{1\\0}\otimes\sqcol{1\\1}
     +\sqrt{\frac{\epsilon}{2}}\sqcol{0\\1}\otimes\sqcol{1\\-1}\nonumber\\
   &=&\frac{1}{\sqrt{2}}\sqcol{\sqrt{1-\epsilon}\\ \sqrt{\epsilon}}
                 \otimes\sqcol{1\\0}
      +\frac{1}{\sqrt{2}}\sqcol{\sqrt{1-\epsilon}\\ -\sqrt{\epsilon}}
                 \otimes\sqcol{0\\1}\,,\qquad
\end{eqnarray}
where we recognize the wave functions of \Eq{6-2} in the idler factors.
This is the polarization-entangled idler-signal state emitted by the source.

\begin{figure}
  \centering
  \includegraphics{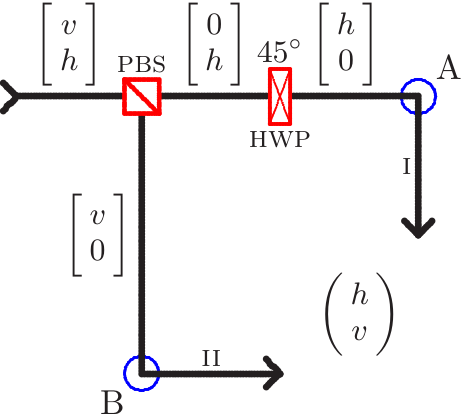}
  \caption{\label{fig:convert}%
    Conversion of a polarization qubit into a path qubit.  
    The polarizing beam splitter (\textsc{pbs}) reflects vertical polarization
    and transmits horizontal polarization; the half-wave plate (\textsc{hwp}) 
    at $45^{\circ}$ turns the horizontal polarization into vertical
    polarization.  
    The overall effect is the conversion of the polarization qubit carried by
    the incoming photon into a path qubit of a vertically polarized photon
    such that the amplitude for the initial horizontal polarization becomes
    the amplitude for path \Path{i}, and the initial vertical-polarization
    amplitude  becomes the path-\Path{ii} amplitude.  
    }
\end{figure}

The signal enters the Mach--Zehnder interferometer loop, which has a
polarization-to-path converter (\textsc{ppc}) at the entrance.
As explained in the caption to \Fig{convert}, it turns the polarization qubit
with the wave function 
\mbox{\small$\Bigl[\begin{array}{c}v\\[-1ex]h\end{array}\Bigr]$}
into a path qubit with the wave function
\mbox{\small$\Bigl(\begin{array}{c}h\\[-1ex]v\end{array}\Bigr)$}.
Accordingly, the idler-signal pair of photons is then prepared in the state
\begin{equation}\label{eq:7-4}
  \frac{1}{\sqrt{2}}\sqcol{\sqrt{1-\epsilon}\\ -\sqrt{\epsilon}}
                 \otimes\column{1\\0}
      +\frac{1}{\sqrt{2}}\sqcol{\sqrt{1-\epsilon}\\ \sqrt{\epsilon}}
                 \otimes\column{0\\1}\,,
\end{equation}
which is exactly the situation depicted in \Fig{same-photon} after the photon
has passed checkpoints A or B, except that now the polarization is that of
the partner photon (idler) rather than the Mach--Zehnder-interfering photon
(signal). 
The further fate of the signal is either as in \Fig{2paths}(a) when the
interferometer is closed and we condition on detecting the photon at
exit~\Path{ii}, or as in \Fig{2paths}(b) when beam splitter \BS{3} is
removed and the photon can be detected at exit~\Path{i} or at 
exit~\Path{ii}. 
With a suitably long delay line in the signal path from the source to the
interferometer, it is possible to postpone Alice's decision between the ``wave
mode'' of \Fig{2paths}(a) and the ``particle mode'' of \Fig{2paths}(b) until
after Bob has measured the idler polarization and announced whether he bets on
one of the signal paths or not.

\begin{figure}
  \centering
  \includegraphics{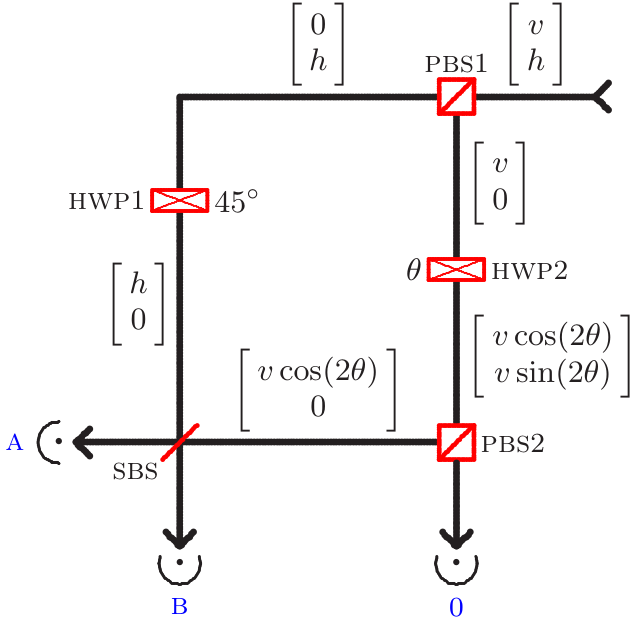}
  \caption{\label{fig:MUD}%
    Unambiguous discrimination of two polarization states.  
    At the entrance, a polarizing beam splitter (\textsc{pbs}{\footnotesize1})
    separates the amplitude $h$ for horizontal polarization from the amplitude
    $v$ for vertical polarization. 
    The horizontally polarized component is converted to vertical polarization
    by a half-wave plate at $45^\circ$ (\textsc{hwp}{\footnotesize1}) and then
    reaches the symmetric beam splitter \textsc{sbs}. 
    It is re-united there with the vertical-polarization amplitude $v$,
    diminished in strength by the combined effect of
    \textsc{hwp}{\footnotesize2}, set at angle $\theta$, and
    \textsc{pbs}{\footnotesize2}.  
    The horizontally polarized amplitude $v\sin(2\theta)$ introduced by
    \textsc{hwp}{\footnotesize2}
    is transmitted by \textsc{pbs}{\footnotesize2} and may then trigger a
    detector (outcome 0). 
    The two other detectors (outcomes A and B) respond to photons at the exit
    ports of \textsc{sbs}.   
    }
\end{figure}

The fate of the idler is different. 
We examine its polarization with the measurement for unambiguous discrimination
(\textsc{mud}) with the outcome operators of \Eq{6-5} for the conclusive
results and the outcome operator of \Eq{6-6} for the inconclusive result.
This measurement can be implemented by the setup of \Fig{MUD}; see
\cite{Clarke+3:01} or \cite{Neves+5:09} for similar but different setups.

An idler  with initial polarization state 
\mbox{\small$\Bigl[\begin{array}{c}v\\[-1ex]h\end{array}\Bigr]$}
is detected by the three detectors with the probabilities
\begin{eqnarray}\label{eq:7-5}
  p^{\ }_{\textsc{a}}&=&\frac{1}{2}\bigl|v\cos(2\theta)-h\bigr|^2
                \quad\mbox{for outcome A}\,,\nonumber\\
  p^{\ }_{\textsc{b}}&=&\frac{1}{2}\bigl|v\cos(2\theta)+h\bigr|^2
                \quad\mbox{for outcome B}\,,\nonumber\\
  p^{\ }_0&=&\bigl|v\sin(2\theta)|^2\quad\mbox{for outcome 0}\,,
\end{eqnarray}
where $\theta$ is the angle at which the half-wave plate \textsc{hwp}{\small2}
in \Fig{MUD} is set.  
Both polarization states to be distinguished have ${v=\sqrt{1-\epsilon}}$, and
there is ${h=-\sqrt{\epsilon}}$ for signal path~\Path{i} (checkpoint A) or 
${h=\sqrt{\epsilon}}$ for signal path~\Path{ii} (checkpoint B). 
Therefore, the choice ${\cos(2\theta)=\sqrt{\epsilon/(1-\epsilon)}}$ ensures
unambiguous discrimination: 
\begin{eqnarray}\label{eq:7-6}
  p^{\ }_{\textsc{a}}&=&\left\{
    \begin{array}{cl}
      2\epsilon & \mbox{for checkpoint A,}\\
         0      & \mbox{for checkpoint B,}
    \end{array}\right.\nonumber\\
  p^{\ }_{\textsc{b}}&=&\left\{
    \begin{array}{cl}
         0      & \mbox{for checkpoint A,}\\
      2\epsilon & \mbox{for checkpoint B,}
    \end{array}\right.\nonumber\\
  p^{\ }_0&=&1-2\epsilon\mbox{\ for both checkpoints.}
\end{eqnarray}

\begin{figure}
  \centering
  \includegraphics{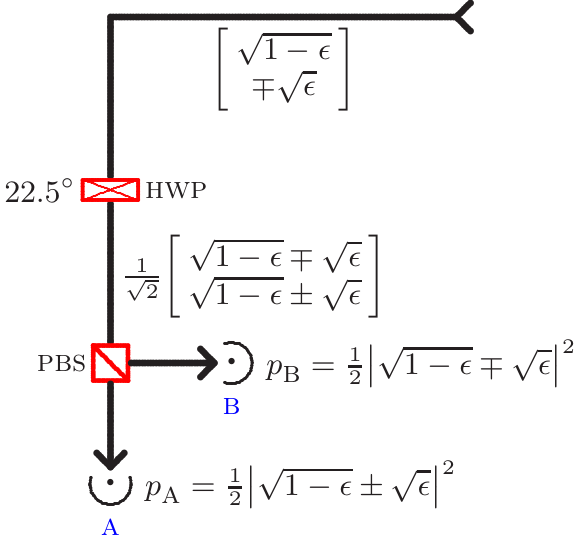}
  \caption{\label{fig:MEM}%
    Measurement for error minimization. 
    The idler passes through the half-wave plate (\textsc{hwp}) 
    at $22.5^{\circ}$ and is then sent to one of the two detectors by the
    polarizing beam splitter (\textsc{pbs}). 
    We bet on signal path~\Path{i} (checkpoint A) if detector~A registers
    the idler, and on path~\Path{ii} (checkpoint B) if detector~B fires.}
\end{figure}

Prior to performing the experiment of \Fig{photonpair}, one should verify 
that the setup is properly aligned.
For this purpose, we vary the phase in the Mach--Zehnder interferometer and
record the interference pattern to confirm that the fringe visibility is
${1-2\epsilon}$. 
We also measure the idler polarization with the error-minimizing measurement
\cite{Helstrom:76} of \Fig{MEM} that maximizes our chance of guessing the path
right \cite{Jaeger+2:95,Englert:96} and so confirm that we guess right for the
fraction $\frac{1}{2}+\sqrt{\epsilon(1-\epsilon)}$ of all signals. 

The experiment of \Fig{same-photon}, in the version of \Fig[Figs.~]{photonpair}
to \Fig[]{MEM}, has been performed and the results are in full agreement with
the theoretical predictions \cite{Len+3:16}.
In particular, the experiment confirms that every signal detected at  
exit~\Path{ii} has a known path through the interferometer irrespective of the
value of $\epsilon$.

\section{Single-photon version of the experiment by Danan %
            \textit{et al.\/} \cite{Danan+3:13}}%
\begin{figure}
  \centering
  \includegraphics{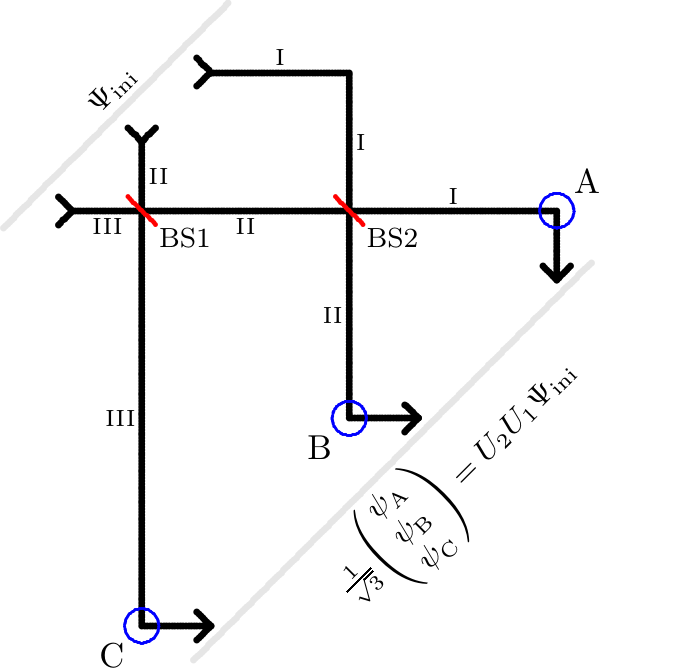}
  \caption{\label{fig:3paths-ini}%
    The preparation stage of Vaidman's interferometer produces an
    equal-weight superposition of the particle's three paths and entangles the
    particle with the path marker.
    The entangled particle--path-marker state of \Eq{8-1}, which describes
    the situation after the particle has passed  checkpoints A, B, or C,
    can be realized by the action of beam splitters \BS{1} and \BS{2} 
    on the pre-entangled state $\Psi_{\mathrm{ini}}$ of \Eq{8-2}.}  
\end{figure}

\label{sec:1photon-3paths}
In the three-path interferometer of \Fig[Figs.~]{3paths}, \Fig[]{fwd-bwd},
\Fig[]{p-mode.a}, and \Fig[]{p-mode.b}, we have the wave function in
\Eq{4-16} for the entangled state of the interfering particle and the path
marker after unitary operators $A$, 
$B$, $C$ acted when the particle passed checkpoints A, B, C and before the
path amplitudes are processed by beam splitters \BS{3} and \BS{4}.
That is, \Eq{8-1} refers to the state of affairs depicted in \Fig{p-mode.a}
before the particle is detected at one of the exit ports or the path marker is
measured for unambiguous path discrimination.
Now, just as we did in Sec.~\ref{sec:signal-idler} for the two-path
interferometer of \Fig[Figs.~]{2paths}--\Fig[]{convert}, we can also here ask:
Which pre-entangled initial state $\Psi_{\mathrm{ini}}$ for the path marker
and the particle would yield \Eq{8-1} upon the action of $U_2U_1$ on the
particle's path amplitudes? 
The answer is
\begin{eqnarray}\label{eq:8-2}
  \Psi_{\mathrm{ini}}&=&\bigl(U_2U_1\bigr)\adj
       \frac{1}{\sqrt{3}}\column{\wf{A} \\ \wf{B} \\ \wf{C}}
\nonumber\\&=&
    \frac{1}{\sqrt{18}}\column{-\sqrt{3}(\wf{B}-\wf{A}) \\
                               2\wf{C}-\wf{B}-\wf{A} \\
                               \sqrt{2}(\wf{C}+\wf{B}+\wf{A})}
\nonumber\\&=&
    \sqrt{\epsilon}\sqcol{1\\0\\0}\otimes\column{1\\0\\0}
   +\sqrt{\epsilon}\sqcol{0\\1\\0}\otimes\column{0\\1\\0}
\nonumber\\&&\mbox{}
   +\sqrt{1-2\epsilon}\sqcol{0\\0\\1}\otimes\column{0\\0\\1}\,,
\end{eqnarray}
as illustrated by \Fig{3paths-ini}.

This observation suggests the following proposal for a single-photon version
of the experiment by Danan \textit{et al.\/} \cite{Danan+3:13},
an analog of the signal-idler scheme of Sec.~\ref{sec:signal-idler}.
There, we have the path qubit of the signal entangled with the
polarization qubit of the idler; here, the path qutrit of the signal
is entangled with a path qutrit of the idler.
The signal traverses Vaidman's three-path interferometer, and the
idler is measured either in accordance with \Eq[Eqs.~]{4-6}--\Eq[]{4-11} 
for the purpose of unambiguous discrimination of the signal path or in
accordance with \Eq[Eqs.~]{5-4}--\Eq[]{5-6} for the defense of common sense.

The entangled idler-signal state of \Eq{7-2} uses a single pair of
corresponding propagation directions on the cone of down-converted photons
emitted by the double-crystal that converts photons from the short-wavelength
pump beam into long-wavelength entangled photon pairs \cite{Kwiat+4:99}.
We now use two pairs of corresponding propagation directions and have the
idler-signal wave function
\begin{eqnarray}\label{eq:8-3}
&&\frac{1}{\sqrt{2}}\Biggl[\left(\sqrt{\frac{\epsilon}{1-\epsilon}}
        \sqcol[@{}c@{}]{1\\0}\!\otimes\!\sqcol[@{}c@{}]{1\\0}
       +\sqrt{\frac{1-2\epsilon}{1-\epsilon}}
        \sqcol[@{}c@{}]{0\\1}\!\otimes\!\sqcol[@{}c@{}]{0\\1}
  \right)\!\otimes\psi^{\ }_{\vec{k}_1}\nonumber\\&&\mbox{}
 +\left(\sqrt{\frac{\epsilon}{1-\epsilon}}
        \sqcol[@{}c@{}]{1\\0}\!\otimes\!\sqcol[@{}c@{}]{1\\0}
       +\sqrt{\frac{1-2\epsilon}{1-\epsilon}}
        \sqcol[@{}c@{}]{0\\1}\!\otimes\!\sqcol[@{}c@{}]{0\\1}
  \right)\!\otimes\psi^{\ }_{\vec{k}_2}\Biggr]\,,\nonumber\\&&
\end{eqnarray}
where the tensor products of the two-component columns have the same meaning
as in \Eq{7-2} and the spatial wave functions $\psi^{\ }_{\vec{k}_1}$ and
$\psi^{\ }_{\vec{k}_2}$ refer to the respective propagation directions,
specified by the signal wave vectors $\vec{k}_1$ and $\vec{k}_2$.
The probability amplitudes of $\sqrt{\epsilon/(1-\epsilon)}$ for the
vertical-vertical components and $\sqrt{(1-2\epsilon)/(1-\epsilon)}$ for the
horizontal-horizontal components are adjusted by setting the polarization of
the pump beam accordingly. 

We remove the
$\sqcol[@{}c@{}]{0\\1}\otimes\sqcol[@{}c@{}]{0\\1}%
\otimes\psi^{\ }_{\vec{k}_1}$ component by a polarizer in the first path
of the signal (or in the corresponding idler path) and so reduce the
wave function in \Eq{8-3} to
\begin{eqnarray}\label{eq:8-4}
  &&\sqrt{\epsilon}\sqcol[@{}c@{}]{1\\0}\!\otimes\!\sqcol[@{}c@{}]{1\\0}
  \!\otimes\psi^{\ }_{\vec{k}_1}\nonumber\\&&\mbox{}
  +\left(\sqrt{\epsilon}
        \sqcol[@{}c@{}]{1\\0}\!\otimes\!\sqcol[@{}c@{}]{1\\0}
       +\sqrt{1-2\epsilon}
        \sqcol[@{}c@{}]{0\\1}\!\otimes\!\sqcol[@{}c@{}]{0\\1}
  \right)\!\otimes\psi^{\ }_{\vec{k}_2}\,.\qquad
\end{eqnarray}
The $\psi^{\ }_{\vec{k}_1}$ component has the idler and the signal vertically
polarized, and the same is the case for the $\psi^{\ }_{\vec{k}_2}$
component after converting the polarization qubits to path qubits as in
\Fig{convert}. 
We then have the wave function
\begin{eqnarray}\label{eq:8-5}
  \Psi_{\mathrm{ini}}
  &=&\sqrt{\epsilon}\column{1\\0\\0}\otimes\column{1\\0\\0}
   +\sqrt{\epsilon}\column{0\\1\\0}\otimes\column{0\\1\\0}
\nonumber\\&&\mbox{}
   +\sqrt{1-2\epsilon}\column{0\\0\\1}\otimes\column{0\\0\\1}
\end{eqnarray}
for the initial path-entangled idler-signal state where the column entries are
path-qutrit probability amplitudes.
This $\Psi_{\mathrm{ini}}$ has exactly the structure of the
$\Psi_{\mathrm{ini}}$ in \Eq{8-2}, with the unspecified path-marker qutrit now
identified with the path qutrit of the idler.

\begin{figure}
  \centering
  \includegraphics{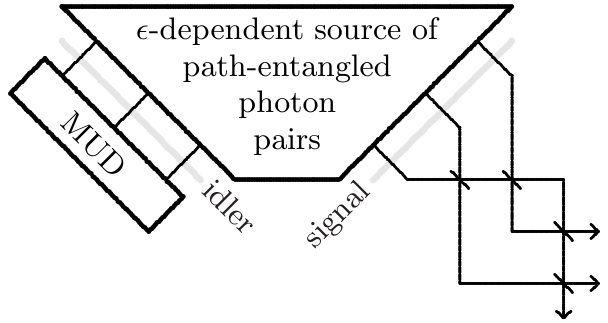}
  \caption{\label{fig:3paths-double}%
    Scheme of the single-photon version of the experiment by Danan \textit{et
      al.} \cite{Danan+3:13}.
    The source prepares idler-signal pairs in the path-entangled state of
    \Eq{8-5}. 
    The signal traverses Vaidman's three-path interferometer of \Fig{3paths}.
    The idler enters the apparatus for the measurement for unambiguous
    discrimination (\textsc{mud}), by which we gain information about the path
    taken by the signal.} 
\end{figure}

The source in \Fig{3paths-double} prepares the idler-signal pairs in the
path-entangled state of \Eq{8-5}.
The signal is fed into Vaidman's three-path interferometer, operated
with two, three, or all four beam splitters as depicted in \Fig{p-mode.a},
\Fig{p-mode.b}, or \Fig{3paths}, respectively.
The idler is measured for unambiguous discrimination of the three
signal paths identified in \Fig{p-mode.a}, or for unambiguous
discrimination of exits \Path{ii} and \Path{iii} in \Fig{p-mode.b}.

\begin{figure}
  \centering
  \includegraphics{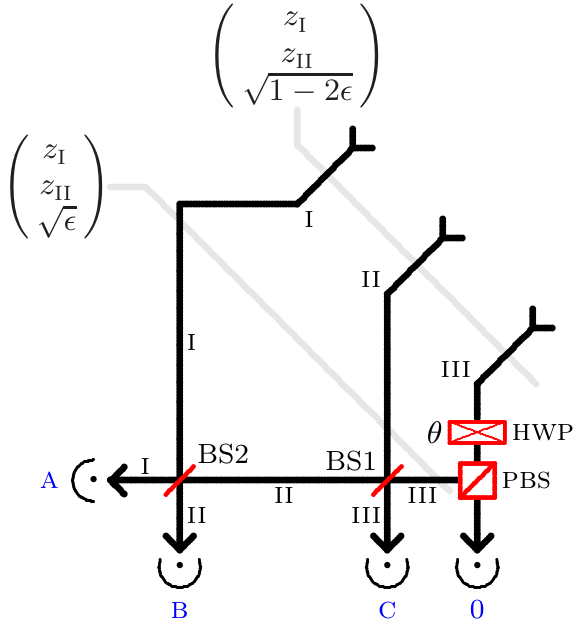}
  \caption{\label{fig:3paths-MUD}%
    Unambiguous discrimination of the three states of the idler path qutrit.
    The arriving idler is vertically polarized.
    A half-wave plate, in conjunction with a polarizing beam splitter, 
    reduces the path-\Path{iii} amplitude from $\sqrt{1-2\epsilon}$
    to $\sqrt{\epsilon}$ before the idler enters a copy of the preparation
    stage of Vaidman's interferometer. 
    The idler is detected after being processed by beam splitters \BS{1}
    and \BS{2} to yield unambiguous information whether the signal passed
    checkpoint A, B, or C.
    The horizontally polarized component introduced by the half-wave plate
    accounts for the inconclusive outcome.} 
\end{figure}

The unambiguous discrimination of the three idler states with the wave
functions of \Eq{4-3} is achieved by the setup of \Fig{3paths-MUD}.
In a first step, a half-wave plate set at angle $\theta$ with
$\cos(2\theta)=\sqrt{\epsilon/(1-2\epsilon)}$ converts the path-\Path{iii} 
amplitude of $\sqrt{1-2\epsilon}$ into amplitude $\sqrt{\epsilon}$ for vertical
polarization and amplitude $\sqrt{1-3\epsilon}$ for horizontal polarization. 
A polarizing beam splitter guides the horizontal component to the ``0''
detector of the inconclusive outcome which, we recall, indicates
path~\Path{iii} for signals detected by D when all four beam splitters are
in place in Vaidman's interferometer. 
The overall effect on the remaining probability amplitudes of the path qutrit
of the vertically polarized idler is the transition
\begin{equation}\label{eq:8-6}
  \column{z^{\ }_{\sPath{i}}\\z^{\ }_{\sPath{ii}}\\\sqrt{1-2\epsilon}} \to
  \column{z^{\ }_{\sPath{i}}\\z^{\ }_{\sPath{ii}}\\\sqrt{\epsilon}}
\end{equation}
where 
\begin{equation}\label{eq:8-7}
  \column{z^{\ }_{\sPath{i}}\\z^{\ }_{\sPath{ii}}}=\left\{
      \begin{array}{c@{\ \mbox{for signal checkpoint\ }}l}
        \column{\sqrt{3\epsilon/2}\\-\sqrt{\epsilon/2}} 
        & \textrm{A,}\\[3ex]
        \column{-\sqrt{3\epsilon/2}\\-\sqrt{\epsilon/2}} 
        & \textrm{B,}\\[3ex]
        \column{0\\\sqrt{2\epsilon}} 
        & \textrm{C.}
      \end{array}\right.
\end{equation}
These three columns of idler-path-qutrit amplitudes are pairwise orthogonal
and make up a three-by-three matrix that is proportional to $(U_2U_1)\adj$,
\begin{equation}\label{eq:8-8}
  \column[ccc]{\sqrt{3\epsilon/2}&-\sqrt{3\epsilon/2}&0\\
               -\sqrt{\epsilon/2}&-\sqrt{\epsilon/2}&\sqrt{2\epsilon}\\
               \sqrt{\epsilon}&\sqrt{\epsilon}&\sqrt{\epsilon}}
               =\sqrt{3\epsilon}\,(U_2U_1)\adj\,.
\end{equation}
Therefore, the application of $U_2U_1$ turns the three columns into
single-path columns,
\begin{equation}\label{eq:8-9}
  U_2U_1\column{z^{\ }_{\sPath{i}}\\z^{\ }_{\sPath{ii}}\\\sqrt{\epsilon}}
   =\left\{\begin{array}{c@{\ \mbox{for signal checkpoint\ }}l}
        \mbox{\small$\column{\sqrt{3\epsilon}\\0\\0}$} 
        & \textrm{A,}\\
        \mbox{\small$\column{0\\\sqrt{3\epsilon}\\0}$}
        & \textrm{B,}\\
        \mbox{\small$\column{0\\0\\\sqrt{3\epsilon}}$}
        & \textrm{C,}\\
      \end{array}\right.
\end{equation}
and this mapping is realized by the preparation-half of Vaidman's
interferometer in \Fig{3paths-MUD}.
Without elaborating on this, we note in passing that the minimum-error
measurement is performed by the setup of \Fig{3paths-MUD} with the half-wave
plate, the polarizing beam splitter, and detector ``0'' removed.

\begin{figure}
  \centering
  \includegraphics{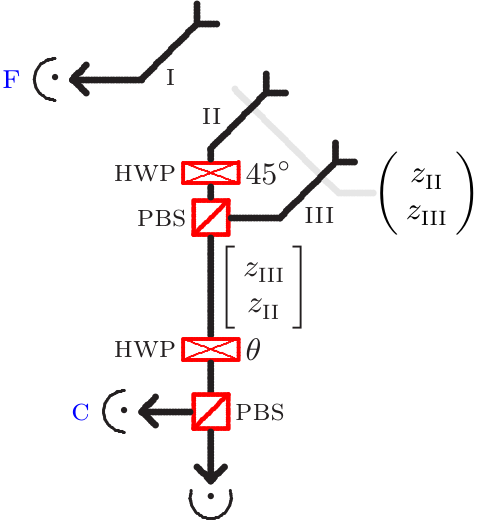}
  \caption{\label{fig:3paths-vN}%
    The orthogonal measurement that distinguishes the idler-path-qutrit states
    of \Eq{5-4}. 
    Upon detection of the idler at  exits ``\textsc{c}'' or
    ``\textsc{f}'', we know that the signal reached beam splitter \BS{4}
    via checkpoints C or F in \Fig{3paths}, respectively, before being
    detected by detector D. 
    No idler is detected at the third exit.} 
\end{figure}

The orthogonal measurement that distinguishes the three idler states with the
wave functions of \Eq{5-4} is realized by the setup of \Fig{3paths-vN}.
For $\phi^{\ }_{\sPath{iii}}$, we simply detect the idler in path~\Path{i}.
For $\phi^{\ }_{\sPath{i}}$ and $\phi^{\ }_{\sPath{ii}}$, we convert the
path qubit of the idler paths \Path{ii} and \Path{iii} with the probability
amplitudes $z^{\ }_{\sPath{ii}}$ and $z^{\ }_{\sPath{iii}}$ into a
polarization qubit (the reversal of the polarization-to-path conversion in
\Fig{convert}), then rotate the polarization by a half-wave plate set at
$\theta$ with $\cos(2\theta)=\sqrt{1-2\epsilon}$ and
$\sin(2\theta)=\sqrt{2\epsilon}$, and finally use a polarizing beam splitter
to separate the resulting vertical and horizontal components.

\section{Summary}
Our analysis of Vaidman's three-path interferometer with weak path marking has
established that common sense does not mislead us.
If there is only one path available for the particle's journey from the source
to the detector, then this path is indeed taken.
Such statements about the actual path through the interferometer are
meaningful only if they represent path knowledge acquired by a suitable
observation.
Here, this is achieved by examining the faint traces left by the particle on
its way from the source to the detector.
Depending on which information we wish to gain, we examine these traces by one
measurement or another.
In the case of Vaidman's interferometer, we can know for each 
particle detected by D whether it arrived via the internal loop or not.
In the limit of ever fainter traces, all detected particles bypass the
loop --- exactly as common sense tells us.
Vaidman's criterion (\ref{eq:2-5}) does not correctly identify the path taken
by the particle. 
These conclusions can be confirmed by a single-particle version of the
experiment by Danan \textit{et al.\/} \cite{Danan+3:13}; we propose
an explicit scheme for that.
Finally, we note that others have also concluded, with a variety of
arguments, that the common sense reasoning is all right, 
notably Li \textit{et al.} \cite{Li+2:13}, Sokolovski \cite{Sokolovski:17a},
and Griffiths \cite{Griffiths:16}.

\begin{acknowledgments}
BGE is sincerely grateful for insightful discussions with Lev Vaidman,
Saverio Pascazio, and Reinhard Werner.
We acknowledge Daniel Terno's valuable feedback on an earlier version of the
text. 
This work is funded by the Singapore Ministry of Education (partly through the
Academic Research Fund Tier 3 MOE2012-T3-1-009) and the National Research
Foundation of Singapore.
HKN is also funded by a Yale-NUS College start-up grant.
\end{acknowledgments}

\onecolumngrid\vspace{\columnsep}

\appendix*
\section{Miscellanea}
For reference, we report some technical details here.
This could be helpful for a reader who wants to confirm various statements in
the text ``on the fly.''

The three ways of splitting the overall unitary operator 
\begin{equation}\label{eq:A-1}  
   U_4U_3U_2U_1=\frac{1}{3}\column[ccc]{0 & -\sqrt 3 & \sqrt 6\\
                                       \sqrt 3 & 2 & \sqrt 2\\
                                       -\sqrt 6 & \sqrt 2 & 1}
\end{equation} 
in the rows of Table~\ref{tbl:weak} are 
\begin{equation}\label{eq:A-2}
  U_4\bigl(U_3U_2U_1\bigl)=
   \frac{1}{\sqrt{3}}\column[ccc]{\sqrt{3} & 0 & 0\\%
                                  0&-1&\sqrt{2}\\%
                                  0&\sqrt{2}&1}
   \frac{1}{\sqrt{3}}\column[ccc]{0 & -1 & \sqrt 2\\
                                  -\sqrt 3 & 0 & 0\\
                                  0 & \sqrt 2 & 1}
\end{equation}
and
\begin{equation}\label{eq:A-3}
   \bigl(U_4U_3\bigr)\bigl(U_2U_1\bigl)= 
    \frac{1}{\sqrt{6}}\column[ccc]{\sqrt 3 & \sqrt 3 & 0\\
                                   1 & -1 & 2\\
                                   -\sqrt 2 & \sqrt 2 & \sqrt 2 }
    \frac{1}{\sqrt{6}}\column[ccc]{\sqrt 3 & -1 & \sqrt 2\\
                                   -\sqrt 3 & -1 & \sqrt 2\\
                                   0 & 2 & \sqrt 2}
\end{equation}
as well as
\begin{equation}\label{eq:A-4}
   \bigl(U_4U_3U_2\bigr)U_1=
   \frac{1}{\sqrt{3}}\column[ccc]{0 & \sqrt 3 & 0\\
                                  1 & 0 & \sqrt 2\\
                                  -\sqrt 2 & 0 & 1}
   \frac{1}{\sqrt{3}}\column[ccc]{\sqrt{3} & 0 & 0\\%
                                  0&-1&\sqrt{2}\\%
                                  0&\sqrt{2}&1}\,.
\end{equation}
Correspondingly, the overall probability amplitude 
\begin{equation}\label{eq:A-5}
  \column{0\\0\\1}\adj U_4U_3U_2U_1\column{0\\0\\1}=\frac{1}{3}
\end{equation}
can be calculated by five different inner products,
\begin{eqnarray}\label{eq:A-6}
  \frac{1}{3}&=&\column{0\\0\\1}\adj\frac{1}{3}\column{\sqrt6\\\sqrt2\\1}
        =\frac{1}{\sqrt3}\column{0\\\sqrt2\\1}\adj
         \frac{1}{\sqrt3}\column{\sqrt2\\0\\1}
        =\frac{1}{\sqrt3}\column{-1\\1\\1}\adj
         \frac{1}{\sqrt3}\column{1\\1\\1}
         =\frac{1}{\sqrt3}\column{-\sqrt2\\0\\1}\adj
         \frac{1}{\sqrt3}\column{0\\\sqrt2\\1}        
\nonumber\\&=&
        \frac{1}{3}\column{-\sqrt6\\\sqrt2\\1}\adj\column{0\\0\\1}\,.
\end{eqnarray}
Each of these products stands for a probability amplitude of the form 
$\langle\textsc{bwd}|\textsc{fwd}\rangle$ between a ``forward-in-time ket''
$|\textsc{fwd}\rangle$ and a ``backward-in-time'' bra $\langle\textsc{bwd}|$.
The blue intensities in \Fig{fwd-bwd} are proportional to the squares of the
amplitudes in the columns that represent the $|\textsc{fwd}\rangle$ kets, and
likewise for the red intensities and the amplitudes of the rows for the
$\langle\textsc{bwd}|$ bras.  
The weak values in Table~\ref{tbl:weak} are the normalized matrix elements
\begin{equation}\label{eq:A-7}
  \frac{\langle\textsc{bwd}|X|\textsc{fwd}\rangle}
       {\langle\textsc{bwd}|\textsc{fwd}\rangle}
       =3\langle\textsc{bwd}|X|\textsc{fwd}\rangle
\quad\text{with}\quad
  X\repr\column[ccc]{1&0&0\\0&0&0\\0&0&0}\quad\text{or}\quad
        \column[ccc]{0&0&0\\0&1&0\\0&0&0}\quad\text{or}\quad
        \column[ccc]{0&0&0\\0&0&0\\0&0&1}
\end{equation}
for paths \Path{i}, \Path{ii}, and \Path{iii}, respectively.

With unitary operators $A$, $B$, $C$, $E$, and $F$ marking the path, the
overall unitary operator is
\begin{equation}\label{eq:A-9}
  U_4\column[ccc]{1&0&0\\0&F&0\\0&0&1}
  U_3\column[ccc]{A&0&0\\0&B&0\\0&0&C}
  U_2\column[ccc]{1&0&0\\0&E&0\\0&0&1}U_1=
\frac{1}{6}\column[ccc]{-3(B-A) & -\sqrt3(B+A)E & \sqrt6(B+A)E\\
\sqrt3F(B+A) & 4C+F(B-A)E & \sqrt2\bigl[2C-F(B-A)E\bigr]\\
-\sqrt6F(B+A) & \sqrt2\bigl[2C-F(B-A)E\bigr] & 2\bigl[C+F(B-A)E\bigr]}\,,
\end{equation}
and the analog of \Eq{A-5},
\begin{equation}
\label{eq:A-10}
\column{0\\0\\1}\adj U_4\column[@{}ccc@{}]{1&0&0\\0&F&0\\0&0&1}
  U_3\column[ccc]{A&0&0\\0&B&0\\0&0&C}
  U_2\column[ccc]{1&0&0\\0&E&0\\0&0&1}U_1\column{0\\0\\1}
= \frac{1}{3}\bigl[C+F(B-A)E\bigr]\,,
\end{equation}
can be calculated by eight different inner products,
\begin{eqnarray}\label{eq:A-11}
 \frac{1}{3}\bigl[C+F(B-A)E\bigr]
&=& \column{0\\0\\1}\tps
    \frac{1}{6}\column[@{}c@{}]{\sqrt6(B+A)E \\
                                \sqrt2\bigl[2C-F(B-A)E\bigr] \\
                                2\bigl[C+F(B-A)E\bigr]}
   =\frac{1}{\sqrt3}\column{0 \\ \sqrt2\\ 1}\tps
    \frac{1}{\sqrt6}\column{(B+A)E \\ F(B-A)E \\ \sqrt2 C}
\nonumber\\&=& 
   \frac{1}{\sqrt3}\column{0 \\ \sqrt2F \\ 1}\tps
   \frac{1}{\sqrt6}\column{(B+A)E \\ (B-A)E \\ \sqrt2 C}
  =\frac{1}{\sqrt3}\column{-F \\ F \\ 1}\tps
   \frac{1}{\sqrt3}\column{AE \\ BE \\ C}
\nonumber\\&=&
   \frac{1}{\sqrt3}\column{-FA \\ FB \\ C}\tps
   \frac{1}{\sqrt3}\column{E \\ E \\ 1} 
  =\frac{1}{\sqrt6}\column{-F(B+A) \\ F(B-A) \\ \sqrt2C}\tps
   \frac{1}{\sqrt3}\column{0 \\ \sqrt2E \\ 1}
\nonumber\\&=&
   \frac{1}{\sqrt6}\column{-F(B+A) \\ F(B-A)E \\ \sqrt2C}\tps
   \frac{1}{\sqrt3}\column{0 \\ \sqrt2 \\ 1}
  =\frac{1}{6}\column{-\sqrt6F(B+A) \\
                      \sqrt2\bigl[2C-F(B-A)E\bigr] \\ 
                      2\bigl[C+F(B-A)E\bigr]}\tps
   \column{0 \\ 0 \\ 1}\,,\qquad
\end{eqnarray}
which one could again read as bra-ket products of the
$\langle\textsc{bwd}|\textsc{fwd}\rangle$ kind only that now the various
column entries are operators acting on the degrees of freedom used for the
path marking, whereas the column entries in \Eq{A-6} are probability
amplitudes for the three paths.
\vspace{\columnsep}\twocolumngrid


\begin{thebibliography}{10}

\bibitem{Vaidman:13a}
L. Vaidman, \pra\ \textbf{87}, 052104 (2013).

\bibitem{Vaidman:14}
L. Vaidman, \pra\ \textbf{89}, 024102 (2014).

\bibitem{Danan+3:13}
A. Danan, D. Farfurnik, S. Bar-Ad, and L. Vaidman,
\prl\ \textbf{111}, 240402 (2013).

\bibitem{Li+2:13}
Z.-H. Li, M. Al-Amri, and Z. Zubairy,
\pra\ \textbf{88}, 046102 (2013).

\bibitem{Vaidman:13b}
L. Vaidman, \pra\ \textbf{88}, 046103 (2013).

\bibitem{Saldanha:14}
P. L. Saldanha, \pra\ \textbf{89}, 033825 (2014).

\bibitem{Salih:15}
H. Salih, Front.\ Phys.\ \textbf{3}, 47 (2015).

\bibitem{Danan+3:15}
A. Danan, D. Farfurnik, S. Bar-Ad, and L. Vaidman,
Front.\ Phys.\ \textbf{3}, 48 (2015).

\bibitem{Svensson:14a}
B. E. Y. Svensson, e-print:arXiv1402.4315 [quant-ph] (2014).

\bibitem{Huang+6:14}
J.-H. Huang, L.-Y. Hu, X.-X. Xu, C.-J. Liu, Q. Guo, H.-L. Zhang, and
S.-Y. Zhu, e-print arXiv:1402.4581 [quant-ph] (2014); 
see also Sec.~3 in \cite{Wu+6:15}. 

\bibitem{Wiesniak:14}
M. Wie\'sniak, e-print arXiv:1407.1739 [quant-ph] (2014).

\bibitem{Svensson:14b}
B. E. Y. Svensson, e-print arXiv:1407.4613 [quant-ph] (2014).

\bibitem{Bartkiewicz+5:15}
K. Bartkiewicz, A. \v{C}ernoch, D. Jav\r{u}rek, K. Lemr, J. Soubusta, and
J. Svol\'ik,
\pra\ \textbf{91}, 012103 (2015).

\bibitem{Wu+6:15}
Z.-Q. Wu, H. Cao, J.-H. Huang, L.-Y. Hu, X.-X. Xu, H.-L. Zhang, and S.Y. Zhu,
Opt.\ Express \textbf{23}, 10032 (2015).

\makeatletter\global\advance\@colroom20pt\relax\set@vsize\makeatother

\bibitem{Alonso+1:15}
M. A. Alonso and A. N. Jordan,
Quantum Stud.: Math.\ Found.\ \textbf{2}, 255 (2015).

\bibitem{Potocek+1:15}
V. Poto\v{c}ek and G. Ferenczi,
\pra\ \textbf{92}, 023829 (2015).

\bibitem{Li+3:15}
F. Li, F. A. Hashmi, J. X. Zhang, and J.-Y. Zhu,
Chin.\ Phys.\ Lett.\ \textbf{32}, 050303 (2015).

\bibitem{Vaidman:16d}
L. Vaidman, \pra\ \textbf{93}, 017801 (2016).

\bibitem{Vaidman:16a}
L. Vaidman, \pra\ \textbf{93}, 036103 (2016).

\bibitem{Bartkiewicz+5:16}
K. Bartkiewicz, A. \v{C}ernoch, D. Jav\r{u}rek, K. Lemr, J. Soubusta, and
J. Svol\'ik,
\pra\ \textbf{93}, 036104 (2016).

\bibitem{Hashmi+3:16}
F. A. Hashmi, F. Li, S.-Y. Zhu, and M. S. Zubairy,
J. Phys.\ A: Math.\ Theor.\ \textbf{49}, 345302 (2016).

\bibitem{Sokolovski:17a}
D.~Sokolovski, \pl\ \textbf{A381}, 227 (2017).

\bibitem{Griffiths:16}
R. B. Griffiths, \pra\ \textbf{94}, 032115 (2016).

\bibitem{Vaidman:16b}
L. Vaidman, e-print arXiv:1610.04781 [quant-ph] (2016).

\bibitem{Vaidman:16c}
L. Vaidman, \pra\ \textbf{95}, 066101 (2017).

\bibitem{Bula+6:16}
M. Bula, K. Bartkiewicz, A. \v{C}ernoch, D. Jav\r{u}rek, K. Lemr, 
V. Mich\'alek, and J. Soubusta,
\pra\ \textbf{94}, 052106 (2016).

\bibitem{Ben-Israel+6:17}
A. Ben-Israel, L. Knips, J. Dziewior, J. Meinecke, A. Danan, H. Weinfurter,
and L. Vaidman, Chin.\ Phys.\ Lett.\ \textbf{34}, 020301 (2017).

\bibitem{Roy+1:17}
A. Roy and S. Ghosh, e-print arXiv:1701.03074 [quant-ph] (2017).

\bibitem{Nikolaev:17a}
G. N. Nikolaev, JETP Lett.\ \textbf{105}, 152 (2017).

\bibitem{Vaidman:17a}
L. Vaidman, JETP Lett.\ \textbf{105}, 473 (2017).

\bibitem{Vaidman:17b}
L. Vaidman, e-print arXiv:1703.03615 [quant-ph] (2017).

\bibitem{Zhou+7:17}
Z.-Q. Zhou, X. Liu, Y. Kedem, J.-M. Cui, Z.-F. Li, Y.-L. Hua, C.-F. Li,
and G.-C. Guo, \pra\ \textbf{95}, 042121 (2017).

\bibitem{Nikolaev:17b}
G. N. Nikolaev, JETP Lett.\ \textbf{105}, 475 (2017).

\bibitem{Griffiths:17}
R. B. Griffiths, \pra\ \textbf{95}, 066102 (2017).

\bibitem{Sokolovski:17b}
D.~Sokolovski, e-print arXiv:1704.02172 (2017).

\bibitem{Aharonov+2:64}
Y. Aharonov, P. G. Bergmann, and J. L. Lebowitz,
Phys.\ Rev.\ \textbf{134}, B1410 (1964).

\bibitem{Aharonov+1:90}
Y. Aharonov and L. Vaidman, \pra\ \textbf{41}, 11 (1990).

\bibitem{Len+3:16}
Y. L. Len, J. Dai, B.-G. Englert, and L. Krivitsky, 
e-print arXiv:1708.01408 [quant-ph] (2017).

\bibitem{Horia:14}
While these particular matrices will serve the purpose,
more general ones can be considered; see, for example,
K. Horia, \textit{Post-selected data in quantum measurements}, 
B. Sc. thesis, National University of Singapore, 2014.

\bibitem{weak1} 
Or is it minus-one particle at checkpoint A and plus-one particle each at
checkpoints B and C? 
See Sec.~16.5 in \cite{Aharonov+1:05}.

\bibitem{note:PowerSpectrum}
Actually, the power spectrum reported in Ref.~\cite{Danan+3:13}, the squared
Fourier transform of an intensity difference, is quadratic in the light
intensity and, therefore, outside the scope of linear optics.

\bibitem{fn:EOM1}
In this respect, the situation is markedly better in the experiment of
Zhou \textit{et al.} \cite{Zhou+7:17}, where electro-optic phase modulators 
at the checkpoints imprint sidebands of several GHz.
A tunable frequency filter selects a narrow frequency range before the photons
are detected by detector D, eventually recording the full spectrum of the
emerging photons.

\bibitem{fn:commute}
This makes the analysis simpler and more transparent but is not absolutely
necessary. 
What really matters is that the three unitary operators $FAE$, $FBE$, and $C$
affect the marker state differently.
Eventually, the extraction of path information relies on
\Eq[Eqs.~]{4-2} and \Eq[]{4-4} and it is not important how they come about.
In assuming independent and uncorrelated degrees of freedom, we follow the
example of Refs.~\cite{Vaidman:13a,Vaidman:14,Danan+3:13} and many of the
papers in the list of Refs.~\cite{Li+2:13} through \cite{Sokolovski:17b}. 
Different authors, however, use different conventions. 
For instance, Griffiths \cite{Griffiths:16} associates one two-state system
(qubit) with each checkpoint.
We do not rely on such a specific choice in Secs.~\ref{sec:WPM},
\ref{sec:WPK1}, and \ref{sec:WPK2}, and the path qutrit of the idler in
Sec.~\ref{sec:1photon-3paths} is not of this kind either.

\bibitem{fn:EOM2}
The same destructive interference suppresses the sidebands associated with
checkpoints E and F in the experiment of Zhou \textit{et al.} \cite{Zhou+7:17}.

\bibitem{Helstrom:76}
C. W. Helstrom, 
\textit{Quantum Detection and Estimation Theory} 
(Academic Press, New York, 1976).

\bibitem{Chefles+1:98}
A. Chefles and S. M. Barnett,
\pl\ \textbf{A250}, 223 (1998).

\bibitem{Peres+1:98}
A. Peres amd D. R. Terno,
J. Phys.\ A: Math.\ Gen.\ \textbf{31}, 7105 (1998).

\bibitem{Clarke+3:01}
R. B. M. Clarke, A. Chefles, S. M. Barnett, and E. Riis, 
\pra\ \textbf{63}, 040305 (2001).

\bibitem{Bergou+2:04}
J. A. Bergou, U. Herzog, and M. Hillery,
Lect.\ Notes Phys.\ \textbf{649}, 417 (2004).

\bibitem{Englert+1:10}
B.-G. Englert and J. \v{R}eh\'a\v{c}ek, \jmo\ \textbf{57}, 218 (2010).

\bibitem{NoClue}
If Bob has no clue, he says ``I cannot infer the path.''
Since a particle is never found in more than one place, he does not
say ``The particle followed all three paths simultaneously.'' 

\bibitem{sorting}
In the jargon of a different context, Bob's sorting of Alice's particles into
subensembles constitutes ``probabilistic remote state preparation.''

\makeatletter\global\advance\@colroom20pt\relax\set@vsize\makeatother

\bibitem{fn:EOM3}
This conclusion is supported by the data reported by Zhou \textit{et al.}
\cite{Zhou+7:17} in theirs Figs.~1(b) and 1(c): When the path through
checkpoint C is blocked, not only the sideband for C is suppressed but also
the carrier frequency that corresponds to the inconclusive outcome of the
unambiguous path discrimination.

\bibitem{weak2}
See Sec.~17.5 in \cite{Aharonov+1:05}.

\bibitem{Kwiat+4:99}
P. G. Kwiat, E. Waks, A. G. White, I. Appelbaum, and P. H. Eberhard,
\pra\ \textbf{60}, R773 (1999).

\bibitem{Dai+4:13}
J. Dai, Y. L. Len, Y. S. Teo, L. A. Krivitsky, and B.-G. Englert,
New J. Phys.\ \textbf{15}, 063011 (2013).

\bibitem{Neves+5:09}
L. Neves, G. Lima, J. Aguirre, F. A. Torres-Ruiz, C. Saavedra, and A Delgado,
New. J. Phys.\ \textbf{11}, 073035 (2009).

\bibitem{Jaeger+2:95}
G. Jaeger, A. Shimony, and L. Vaidman, \pra\ \textbf{51}, 54 (1995).

\bibitem{Englert:96}
B.-G. Englert, \prl\ \textbf{77}, 2154 (1996).

\bibitem{Aharonov+1:05}
Y. Aharonov and D. Rohrlich, 
\textit{Quantum Paradoxes --- Quantum Theory for the Perplexed} 
(Wiley-VCH, Weinheim, 2005).

\end{thebibliography}
\end{document}